\documentclass[fontsize=11pt,a4paper,toc=flat]{arxiv-article}
\usepackage{Definitions}

\usetikzlibrary{calc,arrows.meta}

\usepackage{hyperref}
\graphicspath{{./plots/}{./FIGS/}{./Figures/}}

\preprint{\texttt{NT@UW-26-25}}

\usepackage[most]{tcolorbox}
\usetikzlibrary{arrows.meta,decorations.pathreplacing,positioning}

\newcommand{\vect}[1]{\mathbf{#1}}

\newtcolorbox{explanation}{
  enhanced,
  breakable,
  frame hidden,
  borderline west={1.5pt}{0pt}{CornflowerBlue},
  boxrule=0pt,
  left=8pt,
  right=0pt,
  top=0pt,
  bottom=0pt,
  colback=white
}

\newcommand{\OfficialTitle}{
  Andreev reflections and optimal truncation in the unitary Fermi droplet
}
\title{\setstretch{1.4}
	{\color{Thoughtless}\textls[-20]{\OfficialTitle}}
}

\hypersetup{pdfauthor={Silas R.~Beane, Adèle Le Borgne, Domenico Orlando, Susanne Reffert, Maximilian Schwick},pdftitle={\OfficialTitle},%
	colorlinks=true,linkcolor=ThoughtYouWere,citecolor=ThoughtYouWere,urlcolor=ThoughtYouWere}

\author{%
	\begin{minipage}{.94\textwidth}
		\begin{center} \dosserif%
			{\small
           \textbf{Silas~R.~Beane}\textsuperscript{\ding{74}},
  				\textbf{Adèle~Le~Borgne}\textsuperscript{\ding{73}}, 
				\textbf{Domenico~Orlando}\textsuperscript{\ding{72}\ding{73}},
  				\textbf{Susanne~Reffert}\textsuperscript{\ding{73}}, and
  				\textbf{Maximilian~Schwick}\textsuperscript{\ding{73}} 
			}
		\end{center}
  		\authorBlock{\ding{74}}{\dosserif{}%
        Department of Physics,\\
        University of Washington,\\
        Seattle, WA 98195}
		\authorBlock{\ding{73}}{\dosserif{}%
        Albert Einstein Center for Fundamental Physics,\\
        Institute for Theoretical Physics, University of Bern,\\
        Sidlerstrasse 5, CH-3012 Bern, Switzerland}
      \authorBlock{\ding{72}}{\dosserif{}%
        INFN sezione di Torino.\\
        via Pietro Giuria 1, 10125 Torino, Italy}
	\end{minipage}
}

\date{}

\RequirePackage[a4paper,top=30.6mm,bottom=38.6mm,left=34.6mm,right=34.6mm,footskip=2cm]{geometry}
\numberwithin{equation}{section}

\begin{document}

\begin{titlepage}

  \newgeometry{top=23.1mm,bottom=46.1mm,left=34.6mm,right=34.6mm}

  \maketitle

  \thispagestyle{empty}

  \vfill

  \abstract{\normalfont{}{}
    We study the unitary Fermi gas in an external trap in the double-scaling limit of large flavor number $N$ 
and large charge $Q$ at fixed $Q/N$. At leading order, the problem reduces to the spectral zeta function 
of a \acl{bdg} operator in an inhomogeneous pairing background. The grand potential 
admits a gradient expansion with distinct bulk and edge sectors. We show that when the trap is 
harmonic, the bulk sector expansion terminates: beyond the first three terms every bulk contribution 
integrates to zero. This cancellation is general and follows from zeros of the Euler beta function computed 
in the Son–Wingate effective field theory, independent of the low-energy constants. The truncation persists 
for anisotropic traps in any odd dimension. All higher-order dependence for harmonic potentials is therefore 
carried by the droplet edge. For anharmonic traps, the asymptotic bulk series is factorially divergent. Using a 
heat-kernel \acs{wkb} analysis, we show that its large-order growth, sign oscillations, and optimal truncation are 
fixed by complex periodic orbits of the \acs{bdg} problem, which are analytic continuations of Andreev reflections. 
We numerically confirm the predicted Borel singularities for quartic and sextic traps.
  }
  \vfill

\end{titlepage}

\restoregeometry

\setstretch{1.1}
\tableofcontents
\section{Introduction}\label{sec:intro}

In conformal field theory, sectors of large global charge~$Q$ are
controlled by an effective description of the superfluid state induced
by the charge, and observables like local-operator dimensions organize
themselves in a systematic expansion in inverse powers
of~$Q$~\cite{Hellerman:2015nra,Gaume:2020bmp}. The same logic applies
to nonrelativistic conformal field theories
(\textsc{nrcft}s)~\cite{Favrod:2018xov,Kravec:2018qnu,Kravec:2019djc,Orlando:2020idm,Hellerman:2021qzz,Beane:2024kld}.
There the nonrelativistic state-operator
correspondence~\cite{Nishida:2007pj} identifies the dimension of the
lowest operator of charge~$Q$ with the ground-state energy of $Q$
particles in a harmonic trap, measured in units of the trap
frequency. The classic example is the unitary Fermi gas, whose
low-energy dynamics is largely constrained by nonrelativistic general coordinate
and conformal invariance~\cite{Son:2005rv}. Because the relevant observable is the energy of a trapped
superfluid cloud, the large-charge expansion connects directly with cold-atom
experiments and with \emph{ab initio} calculations of small trapped
clusters~\cite{Carlson:2014pxa,Beane:2025hon}, and, via the proximity
of low-energy nucleon interactions to unitarity, with nuclear
phenomena~\cite{Beane:2025tum}.

In contrast with the relativistic case, the trapped gas has a feature
with far-reaching consequences: the finite-density state is spatially
inhomogeneous. The superfluid fills a droplet whose density vanishes
at a finite radius, and close to this surface the gradient expansion
around the \ac{lda} breaks down. A
complete description therefore combines a bulk expansion with a
separate expansion in a boundary layer at the droplet
edge~\cite{Hellerman:2020eff,Hellerman:2021qzz}. In the \ac{eft}, the bulk terms involve low-energy
constants that are not fixed by symmetry, so the structure of the
expansion---i.e., which terms appear, whether it converges, how it
should be truncated---cannot be settled within the \ac{eft} alone.

\paragraph{Setup.}
A systematic microscopic method for addressing these questions is
provided by the large-$N$ limit. In the double-scaling limit
$N\to\infty$, $Q\to\infty$ with $Q/N$ fixed, the large-charge sector
of vector models is dominated by a semiclassical saddle, and the
leading-order problem reduces to the spectral problem of a
one-particle operator on that
background~\cite{Alvarez-Gaume:2019biu,Dondi:2022zna}. For the unitary
Fermi gas, large-$N$ methods have a long history
\cite{Veillette:2007zz}, and the large-charge, large-$N$ formulation
of Ref.~\cite{Hellerman:2023myh} reduces the leading-order grand
potential to the spectral zeta function of a \ac{bdg} operator. In this setting the coefficients of the large-charge
expansion can be computed to high order, and one can ask about its
large-order behavior. For relativistic theories on the sphere, the
answer is that the expansion is asymptotic and resurgent: its Borel
singularities are fixed by nontrivial saddles of a worldline
representation of the heat kernel, which also determine the optimal
truncation and the size of the exponentially small corrections~\cite{Dondi:2021buw,Bersini:2025lxs}.

In this paper, we carry out the corresponding analysis for the unitary
Fermi gas in a trap at large $N$ and large charge \(Q\). Two features make it
substantially richer than its homogeneous counterpart. First, the trapping
potential breaks translation invariance, so the \ac{bdg} operator cannot be
diagonalized in momentum space and its entries are non-commuting
differential operators. Second, the chemical potential and the pairing gap
vanish together at the droplet edge, which acts as a turning surface for
the bulk gradient expansion.
We will concentrate on the double-scaling limit \(N \to \infty\), \(Q \to \infty\) with \(Q/N = q \) fixed and large \(q \gg1\).
Then, in appropriate units, the large-charge
expansion is a small-gradient expansion in a single parameter
$\epsilon$, with $Q/N\sim\epsilon^{-3}$, and the leading-order grand
potential takes the form
\begin{equation}
  \frac{\Omega}{\mu N} \sim \epsilon^{-3}\sum_{n\geq 0} a_n\,\epsilon^{2n}
  \;+\; (\text{edge}) .
\end{equation}
Here we focus on the bulk coefficients~$a_n$.
The optimal-truncation estimates below concern the bulk expansion.
The edge contributions, which are nonvanishing and require their own boundary-layer analysis, will
be treated in detail in future work.

\paragraph{Results.}
Our first result concerns the harmonic trap. There the bulk expansion of $\Omega$
is not asymptotic but \emph{terminates}: beyond the first three terms,
every bulk contribution to the grand potential integrates to zero. This
cancellation is not an artifact of large~$N$. In the Son--Wingate
\ac{eft}~\cite{Son:2005rv}, the contribution of a generic bulk
operator to the grand potential is proportional to an Euler beta
function, and for $V\propto r^2$ in odd spatial dimension this beta
function vanishes for every term carrying a negative power of the
chemical potential. The statement is therefore independent of the
\ac{eft} low-energy constants, and it persists for anisotropic
harmonic traps.
It is tempting to conjecture that this cancellation extends to the quantum corrections of the \ac{eft}, so that no additional bulk logarithms arise beyond the leading one identified in Ref.~\cite{Hellerman:2021qzz}, because of lack of possible counterterms.
However, there remain two important caveats.
The termination is in the bulk contribution
alone; all higher-order dependence is carried by the edge sector. We
emphasize that the truncation holds for the grand potential at fixed
chemical potential: the Legendre transform to fixed charge reorganizes
the series, so the operator dimension $\Delta(Q)$ does not itself
terminate, but is in general an algebraic function of three parameters.
The same mechanism produces exact zeros in other traps;
for $V\propto r^6$, every bulk coefficient with index divisible by
three vanishes.

\bigskip

Our second result concerns anharmonic traps, where the bulk series is
factorially divergent, $a_n\sim(2n)!/A^{2n}$. Using a heat-kernel
\ac{wkb} analysis, we show that the
saddles of the proper-time integral defining the spectral zeta function
are periodic, zero-energy cycles of the squared \ac{bdg} operator. Since the
system is gapped, there are no real quasiparticle orbits of this kind: at
zero energy the momentum is complex throughout the bulk, and the relevant
trajectories are analytic continuations into a classically forbidden
region, as is the case in ordinary tunneling. The zero-energy spectral curve
factorizes into particle-like and hole-like components, which meet
degenerately at the turning points. Resolving these degeneracies at small
nonzero energy, we classify the closed cycles into inner-sheet exchanges,
momentum reversals, and their composites. The inner-sheet exchanges are
the complex analogue of Andreev reflection~\cite{Andreev:1964}: a
particle-like quasiparticle returns as a hole-like one, the contributions
of the common Fermi momentum cancel, and the resulting action is the
smallest available. For monomial traps $V\propto r^p$, the minimal cycles
connect adjacent complex roots of $r^p=1$, and we give their action in
closed form. Except in the harmonic case, these cycles close in complex
configuration space and do not correspond to the physical diameter of the
cloud. A Euclidean instanton is a classical configuration of a quantum path integral, not necessarily a classical trajectory of a physical particle.

\bigskip

We test these predictions against the bulk coefficients computed
independently from the microscopic expansion, through order
$\epsilon^{44}$ for the quartic trap and $\epsilon^{40}$ for the
sextic trap. Fitting the growth and the sign oscillations of the
coefficients, without using the \ac{wkb} actions as input, we recover
the modulus of the leading Borel singularities,
$A_{\text{fit}}= 1.2754 \pm 0.0017$ versus $A_{\textsc{wkb}}=1.2764$ for
$V=r^4$ and $A_{\text{fit}}=0.94066\pm0.00023$ versus
$A_{\textsc{wkb}}=0.94049$ for $V=r^6$, together with their phases,
which produce the characteristic $+,+,-,-$ and $+,-,0$ sign patterns.
The fits determine the positions of the singularities and the relative
Stokes data; the absolute Stokes constants are measured, not predicted.
In this sense we distinguish three notions: the
cycles of the spectral curve, the elementary semiclassical configurations
they define, and the Borel singularities of the particular observable
under study. The relation between the latter two is mediated by
differences of actions and by Stokes data, and our numerical tests
concern the leading singularities of the bulk grand potential.

\bigskip

Practically, the identification of the leading actions fixes the optimal
truncation of the bulk expansion: retaining $N_{\text{opt}}\simeq
A/(2\epsilon)$ terms gives an accuracy of order $e^{-A/\epsilon}$,
exponentially small in $(Q/N)^{1/3}$. For the harmonic trap the
nontrivial semiclassical geometry survives, but because the bulk
coefficients vanish beyond finite order, the usual relation between
large-order growth and the candidate actions cannot be tested directly.
Only in the harmonic case does the grand potential translate directly
into operator dimensions; for anharmonic traps the results describe
energies of trapped clouds, which are directly relevant for experiments
with non-harmonic confinement~\cite{Beane:2025hon}.

\bigskip
The paper is organized as follows.
In Section~\ref{sec:largeNfermions} we review the large-$N$ formulation of the unitary Fermi gas in a trap and introduce the associated spectral problem.
In Section~\ref{sec:heatkernel} we develop the heat-kernel \ac{wkb} approach, formulate the perturbative bulk expansion and the matching problem at the droplet edge, identify the complex zero-energy cycles and their actions, and describe the numerical analysis of the large-order behavior.
In Section~\ref{sec:Examples} we explicitly study the cases $V=0$, $r^2$, $r^4$ and $r^6$.
We establish the termination of the harmonic bulk expansion using the \ac{eft} and compare the semiclassical predictions for the leading Borel singularities with the perturbative coefficients of the anharmonic traps, determining their optimal truncation.
In Section~\ref{sec:fixed-charge} we discuss the Legendre transform to fixed charge, its implications for the asymptotic expansion, and the relation to operator dimensions in the harmonic trap.
We conclude in Section~\ref{sec:conclusions} and give an outlook on further research directions.
In Appendix~\ref{sec:CubeBall} we examine the cube and the ball as complementary examples illustrating the relation between boundary geometry, perturbative expansions, and exponentially suppressed contributions. 
In Appendix~\ref{sec:numerics}, we list the numerically obtained perturbative coefficients of the leading-large-$N$ bulk grand potential at fixed chemical potential.

\section{Large-N fermions at unitarity}\label{sec:largeNfermions}

We consider the unitary Fermi gas at large $N$ and large charge.
The system has \(N\) flavors and a total of \(Q\) particles distributed among them.
We take the double-scaling limit 
\begin{align}
	N &\to \infty, & Q&\to \infty, & \frac{Q}{N} = q \text{ fixed}.
\end{align}
The general strategy is the same as in the relativistic large-$N$ examples of~\cite{Dondi:2022zna,Dondi:2021buw}:
we first work grand canonically at fixed chemical potential $\mu$, where
the large-$N$ path integral is dominated by a saddle, and subsequently pass to fixed charge via a Legendre transform. 
In the present non-relativistic problem the state-operator correspondence maps the lowest operator of charge $Q$ to the ground state of the gas in a harmonic trap. 
A key difference from the relativistic case, with far-reaching conceptual and technical consequences, is the spatial inhomogeneity of the finite-density saddle. In particular, translation invariance is lost, and the droplet has an edge where the particle density vanishes.

Following~\cite{Hellerman:2023myh}, we consider $N$ flavors of
nonrelativistic fermions with two hyperfine states,
$\psi_{\alpha a}$, where $a=1,\ldots,N$ and
$\alpha=\uparrow,\downarrow$. The particles are confined by a potential \(V(\mathbf{r})\) at chemical potential \(\mu\).\footnote{We set $\hbar=m=1$ throughout.} 
At zero temperature the Euclidean action is given by
\begin{equation}
  \label{eq:fermion-action}
  S[\psi]
  =
  \int \dd{\tau} \dd^3{\mathbf r}
  \bigg[
  \bar\psi_{\alpha a}
  \left(
    \partial_\tau-\frac{\Delta}{2}-\mu + V(\mathbf r)
  \right)
  \psi_{\alpha a}
  +
  \frac{2u_0}{N}
  \bar\psi_{\uparrow a}\bar\psi_{\downarrow a}
  \psi_{\downarrow b}\psi_{\uparrow b}
  \bigg],
\end{equation}
The attractive contact coupling $u_0<0$ is tuned to the unitary point,
where the $s$-wave scattering length diverges and all intrinsic scales disappear, allowing a description in terms of a \ac{nrcft}.
In the spirit of the \ac{lda} we define the local chemical potential
\begin{equation}
  \label{eq:local-chemical-potential}
  \mu(\mathbf r)=\mu-V(\mathbf r).
\end{equation}
As in the standard large-$N$ treatment of four-fermion theories, we decouple 
the interaction in the Cooper channel by introducing a complex Hubbard--Stratonovich field: 
\begin{align}
\sigma(\tau,\mathbf r)
&=
-\frac{2u_0}{N}
\sum_{a=1}^N
\psi_{\downarrow a}(\tau,\mathbf r)
\psi_{\uparrow a}(\tau,\mathbf r).
\label{eq:HS-field}
\end{align}
In terms of the Nambu spinor
\begin{align}
\Psi_a
&=
\begin{pmatrix}
\psi_{\uparrow a}\\
\bar\psi_{\downarrow a}
\end{pmatrix},
&
\bar\Psi_a
&=
\Psi_a^\dagger ,
\end{align}
the fermions appear quadratically and can be integrated out. The
resulting effective action is
\begin{equation}
S_{\text{eff}}[\sigma]
=
-N\,\Tr\log\left|G^{-1}[\sigma]\right|
-\frac{N}{2u_0}
\int d\tau\,d^3\mathbf r\,|\sigma|^2 .
\label{eq:sigma-effective-action}
\end{equation}
The overall factor of $N$ makes the saddle-point approximation exact at
leading order in the large-$N$ expansion. We write
\begin{align}
\sigma(\tau,\mathbf r)
&=
\sigma_0(\mathbf r)
+\frac{1}{\sqrt N}\,\widehat\sigma(\tau,\mathbf r),
&
\left.
\frac{\delta S_{\text{eff}}}
{\delta\sigma(\tau,\mathbf r)}
\right|_{\sigma=\sigma_0}
&=
0 .
\label{eq:large-N-gap}
\end{align}
The saddle $\sigma_0(\mathbf r)$ is the position-dependent pairing gap. In the following, we denote this static saddle profile simply by $\sigma(\mathbf r)\equiv\sigma_0(\mathbf r)$.

For a static saddle, the inverse fermion propagator can be written in
terms of the \ac{bdg} operator $B$ as
\begin{align}
G_0^{-1}
&=
-\mathbf 1_2\,\partial_\tau+B(\mathbf r),
&
B(\mathbf r)
&=
\begin{pmatrix}
-h(\mathbf r) & \sigma(\mathbf r)\\
\sigma^*(\mathbf r) & h(\mathbf r)
\end{pmatrix},
\label{eq:BdG-operator}
\\
h(\mathbf r)
&=
-\frac{\Delta}{2}+V(\mathbf r)-\mu .
\label{eq:one-particle-hamiltonian}
\end{align}
At zero temperature the Matsubara sum reduces the fermionic determinant
to the spectral problem of $B$,
\begin{align}
\lim_{\beta\to\infty}
\Tr\log\left|G_0^{-1}\right|
&=
\frac{\beta}{2}\Tr|B|.
\label{eq:det-to-zeta}
\end{align}
At the static saddle, the zero-temperature grand potential
is obtained from the effective action per unit Euclidean time:
\begin{equation}
  \Omega(\mu)
  =
  \lim_{\beta\to\infty}
  \frac{1}{\beta}
  \eval*{S_{\text{eff}}}_{\sigma=\sigma_0}.
\end{equation}
Its Legendre transform gives the free energy,
\begin{align}
  \label{eq:Legendre-transform}
  Q
  &=
    -\frac{\partial\Omega}{\partial\mu},
  &
    F(Q)
  &=
    \Omega(\mu(Q))+\mu(Q)Q .
\end{align}
In a critical theory in a harmonic trap, the non-relativistic state-operator correspondence identifies this value with the dimension of the lowest operator of charge \(Q\):
\begin{align}
\Delta(Q)
&=
\frac{F(Q)}{\omega}.
\label{eq:state-operator}
\end{align}
This is the non-relativistic counterpart of the large-$N$ fixed-charge construction used in~\cite{Dondi:2021buw,Dondi:2022zna}.

It is convenient to work with the positive operator $B^2$.
Explicitly,
\begin{align}
B^2
&=
\begin{pmatrix}
h^2+|\sigma|^2 & [\sigma,h]\\
[h,\sigma^*] & h^2+|\sigma|^2
\end{pmatrix}.
\label{eq:B2-general}
\end{align}
In particular,
\begin{equation}
\Tr|B| = \zeta_{B^2}\lrp*{-\tfrac{1}{2}} .
\end{equation}
with
\begin{equation}
  \zeta_{B^2}(s)  = \Tr(B^{-2s})\,.
\end{equation}
In this regularization, the unitary point corresponds to \(u_0 \to \infty\) and 
the leading large-$N$ grand potential is determined by $\zeta_{B^2}(-1/2)$.
We use the normal-ordering convention in which the empty nonrelativistic vacuum has zero energy, independently of the trapping potential.
This is also consistent with the state–operator correspondence, which identifies the zero-charge state with the identity operator.

The form of \(B^2\) already reveals a key difference from the translation-invariant problem. 
In the trapped problem neither $V(\mathbf r)$ nor $\sigma(\mathbf r)$ is constant, and the entries of $B^2$ are non-commuting differential operators.%
\footnote{With a rotation of the \(2 \times 2\) matrix, the operator \(B^2\) can be recast in the form
\begin{equation}
  B^2 =
  \begin{pmatrix}
    D_+ D_- & 0 \\
    0 &D_- D_+
  \end{pmatrix}
\end{equation}
where \(D_{\pm} = h \pm i \sigma \) (for real \(\sigma\)), which is reminiscent of super quantum mechanics.
This basis is useful for computations, but we will \emph{not } use it in the following because it makes the particle/hole interpretation less transparent.}

In this work we will mainly concentrate on the large double-scaling parameter \(q \gg 1\) limit.
Then it is convenient to write the spectral zeta function as a Mellin integral, 
\begin{align}
\zeta_{B^2}(s)
&=
\frac{1}{\Gamma(s)}
\int_0^\infty\frac{d\tau}{\tau}\,
\tau^s
\int d^3\mathbf r\,
\Tr K_{B^2}(\mathbf r,\mathbf r;\tau),
\label{eq:B2-zeta}
\end{align}
where the heat kernel $K_{B^2}$ is defined by
\begin{align}
\left(
\partial_\tau+B^2(\mathbf r_1)
\right)
K_{B^2}(\mathbf r_1,\mathbf r_2;\tau)
&=
0,
\label{eq:heat-equation}
\\
\lim_{\tau\to0^+}
K_{B^2}(\mathbf r_1,\mathbf r_2;\tau)
&=
\delta(\mathbf r_1-\mathbf r_2)\,
\mathbf 1_2 \,.
\label{eq:heat-initial-condition}
\end{align}
We now introduce the dimensionless variables that will be used in the rest of the paper. For the harmonic potential
\begin{align}
V_{\text{phys}}(\mathbf r)
&=
\frac{\omega^2}{2} r^2,
&
R_{\text{cl}}
&=
\frac{\sqrt{2\mu}}{\omega},
\end{align}
where $R_{\text{cl}}$ is the classical radius of the cloud, \emph{i.e.} the distance from the origin to where the leading-order density vanishes. We rescale
lengths and energies according to
\begin{align}
\mathbf r_{\text{phys}}
&=
R_{\text{cl}}\,\bar{\mathbf{r}},
&
B_{\text{phys}}
&=
\mu\,\bar{{B}},
&
\sigma_{\text{phys}}
&=
\mu\,\bar{\sigma},
&
V_{\text{phys}}
&=
\mu\, \bar{V} .
\label{eq:dimensionless-variables}
\end{align}
From now on we omit the bars that would otherwise distinguish the dimensionless quantities.

Since we are at \(q \gg 1 \), our natural semiclassical expansion parameter is
\begin{align}
\epsilon
&=
\frac{1}{\sqrt{2\mu}\,R_{\text{cl}}}
=
\frac{\omega}{2\mu},
\label{eq:epsilon}
\end{align}
and the dimensionless one-particle operator becomes
\begin{align}
h(\mathbf r)
&=
-\epsilon^2\Delta - \mu(\mathbf r),
&
\mu(\mathbf r)
&=
1 - V(\mathbf r) ,
\label{eq:dimensionless-h}
\end{align}
which shows explicitly how the large-charge expansion in this case coincides with a small-gradient expansion.
The combination \(\mu(\mathbf{r}) = 1 - V(\mathbf{r})\) plays the role of a ``position dependent chemical potential'', at least in the leading-\(\epsilon\) approximation (which is equivalent to the \ac{lda}).

In rescaled coordinates, the harmonic trap is
\begin{align}
V(\mathbf r)
&=
r^2,
&
\mu(\mathbf r)
&=
1 - r^2 .
\label{eq:harmonic-W}
\end{align}
At leading order, the cloud occupies the sphere $r<1$.
In the bulk the leading large-charge gap profile is
\begin{align}
\sigma(\mathbf r)
&=
y_0 \pqty*{1-r^2} + \order{\epsilon^2} ,
&
y_0
&\simeq
1.1622 .
\label{eq:leading-gap}
\end{align}
In the double-scaling large-$N$, large-charge regime one has $Q/N\sim\epsilon^{-3}$, so the large-charge expansion is equivalently a semiclassical expansion in $\epsilon$.

There are two related reasons why the resulting spectral problem is substantially more involved than its translation-invariant counterpart.
The first is that the trapping potential breaks translation invariance, so $B^2$ cannot be diagonalized globally in momentum space.
Second, both chemical potential and gap vanish at the edge \(r = 1\):
\begin{align}
  \mu(r)
  &\xrightarrow[r \to 1]{} 0,
  &
    \sigma(r)
  &\xrightarrow[r \to 1]{} 0 .
\end{align}
The bulk gradient expansion is  not uniform: the edge is a turning-point region requiring a separate boundary-layer description.
This distinction between a slowly varying bulk and a singular edge is central to the analysis below.

The heat-kernel formulation also makes contact with the resurgent large-charge analysis of~\cite{Dondi:2021buw,Bersini:2025lxs}. 
There, the small-proper-time expansion of the heat trace gives the perturbative large-charge series, while exponentially small corrections are associated with non-trivial saddles of a worldline path integral and
control its large-order behavior. Our aim is to determine the analogous structure for the inhomogeneous unitary Fermi gas. Schematically, we seek a trans-series completion of the form
\begin{align}
\zeta_{B^2}(s;\epsilon)
&\sim
\zeta_{B^2}^{(0)}(s;\epsilon)
+
\sum_\gamma
e^{-S_\gamma/\epsilon}\,
\zeta_{B^2}^{(\gamma)}(s;\epsilon),
\label{eq:zeta-transseries}
\end{align}
where $\zeta_{B^2}^{(0)}$ denotes the perturbative heat-kernel sector and the labels $\gamma$ denote non-trivial semiclassical trajectories. The $\zeta_{B^2}^{(\gamma)}(s;\epsilon)$ should themselves be understood as perturbative expansions in $\epsilon$.
The actions $S_\gamma$ determine the nonperturbative scales and, through the associated Borel singularities, the large-order growth and optimal truncation of the perturbative expansion.

\section{Heat Kernel WKB}\label{sec:heatkernel}
\subsection{Rescaling}\label{sec:general-framework}

As explained in Section~\ref{sec:largeNfermions}, the thermodynamics reduces to a spectral analysis of $\left|B\right|$. It is more convenient, however, to study the heat kernel of $B^2$.
It contains the same spectral information and satisfies a differential equation, making it more tractable.

We want to solve the heat kernel problem%
\begin{equation} 
\label{eq:HeatK_Problem_DimLess}
\begin{dcases}
     \left( \del_{\tau}{} +  B^2(\vect{r}_1) \right) K_{B^2}(\vect{r}_1, \vect{r}_2; \tau) = 0, \\
    \lim_{\tau\to0^+} K_{ B^2}(\vect{r}_1, \vect{r}_2; \tau) = \delta(\vect{r}_1 - \vect{r}_2) \cdot \Id_2 .
\end{dcases}
\end{equation}
The rescaling in Eq.~\eqref{eq:dimensionless-variables} is suited to a perturbative treatment.
A further rescaling instead leads to a \ac{wkb} treatment:
\begin{equation}
t = \epsilon \tau.
\end{equation}
The heat-kernel equation then becomes
\begin{equation} 
\label{eq:HeatK_Problem_WKB}
\begin{dcases}
     \left( \epsilon\del_{t}{} +  B^2(\vect{r}_1) \right) K_{ B^2}(\vect{r}_1, \vect{r}_2; t) = 0, \\
    \lim_{t\to0^+} K_{ B^2}(\vect{r}_1, \vect{r}_2; t) = \delta(\vect{r}_1 - \vect{r}_2) \cdot \Id_2 .
\end{dcases}
\end{equation}
We can decompose the heat-kernel equation as follows:
\begin{equation}
\label{eq:HeatK_Problem_WKB_decomposed}
\left(\epsilon\partial_t\Id_2+
\left(h(\vect{r})^2+\sigma(\vect{r})^2\right)\Id_2-
\begin{pmatrix}
0&[\sigma(\vect{r}),\Delta]\\
[\Delta,\sigma(\vect{r})^*]&0
\end{pmatrix}\epsilon^2\right)
K_{ B^2}(\vect{r}_1,\vect{r}_2;t)=0.
\end{equation}
The \ac{wkb} ansatz takes the form
\begin{equation}\label{eq:WKB-ansatz}
  K_{B^2}(\vect{r}_1,\vect{r}_2;t)=\sum_\gamma
  e^{-\frac{\phi_\gamma(\vect{r}_1,\vect{r}_2;t)}{\epsilon}}
  \mathbb{P}_{\gamma, \epsilon }(\vect{r}_1,\vect{r}_2;t),
\end{equation}
where the sum runs over semiclassical trajectories,
including the perturbative sector and nonperturbative exponential corrections with their associated perturbative expansions. The amplitudes $\mathbb{P}_{\gamma,\epsilon}$ are \(2 \times 2\) matrices. Each trajectory contributes its own exponential action and perturbative amplitude to the heat kernel. The perturbative expansion is more conveniently developed using the original $\tau$ scaling, so we treat the two expansions separately. 
\subsection{Perturbative analysis}\label{sec:perturbative}
In this section, we review how the perturbative part of the grand potential is computed. The gradient expansion breaks down at the droplet edge where it must be matched to a local Airy problem~\cite{Kravec:2018qnu,Hellerman:2020eff,Hellerman:2023myh}.

We begin with the perturbative sector, using the dimensionless variables of Eq.~\eqref{eq:HeatK_Problem_DimLess}.
For $\tau\to 0^{+}$, $K(\vect{r}_1,\vect{r}_2 ; \tau)$ must satisfy
\begin{equation}
K(\vect{r}_1,\vect{r}_2 ; \tau)\xrightarrow{\tau\to 0^{+}}\delta(\vect{r}_1 - \vect{r}_2) \cdot \Id_2.
\end{equation}
Because the initial condition is distributional, we work in Fourier space and perform a Wigner--Weyl transform:
\begin{equation}
K_{\text{pert}}(\vect{r}_1,\vect{r}_2 ; \tau)
=
\int \frac{\dd^d p}{(2\pi\epsilon)^d}
\exp*(
\frac{i}{\epsilon}
\vect{p}\cdot(\vect{r}_1-\vect{r}_2))
\mathcal{K}_{\text{pert}}\lrp*{
\tau,\frac{\vect{r}_1+\vect{r}_2}{2},
\vect{p};\epsilon} ,
\end{equation}
where we introduced the symbol \(\mathcal{K}_{\text{pert}}\).

\begin{figure}
  \begin{center}
    \begin{tikzpicture}[
      x=2.5cm,
      y=1cm,
      font=\small
      ]
      \definecolor{bulkblue}{RGB}{45,100,150}
      \definecolor{outergray}{RGB}{100,110,120}

      \fill[outergray!9] (-2.15,0.08) rectangle (-1,1.35);
      \fill[bulkblue!10] (-1,0.08) rectangle (1,1.35);
      \fill[outergray!9] (1,0.08) rectangle (2.15,1.35);

      \node[text=outergray] at (-1.575,1.08) {outside};
      \node[text=bulkblue]  at (0,1.08)      {bulk};
      \node[text=outergray] at (1.575,1.08)  {outside};

      \node at (-1.575,0.58) {$p=0$};
      \node[text=bulkblue] at (0,0.58)
      {$p^2=\mu(r)=1-r^2$};
      \node at (1.575,0.58) {$p=0$};
      \node[text=outergray] at (-1.575,0.25) {\textcircled{2}};
      \node[text=bulkblue]  at (0,0.25)      {\textcircled{1}};
      \node[text=outergray] at (1.575,0.25)  {\textcircled{2}};
      \draw[densely dashed,outergray!70]
      (-1,-0.12) -- (-1,1.48);
      \draw[densely dashed,outergray!70]
      (1,-0.12) -- (1,1.48);

      \draw[->,thick] (-2.3,0) -- (2.35,0)
      node[right] {$r$};

      \foreach \x/\lab in {-1/-1,0/0,1/1}{
        \draw[thick] (\x,0.055) -- (\x,-0.055);
        \node[below=5pt] at (\x,0) {$\lab$};
      }

      \fill[bulkblue] (-1,0) circle (2pt);
      \fill[bulkblue] (1,0) circle (2pt);

    \end{tikzpicture}
  \end{center}
  \caption{The perturbative solution for the harmonic example.
    It automatically splits into bulk and exterior solutions when studying the momentum trace.}
  \label{fig:inside-outside}
\end{figure}
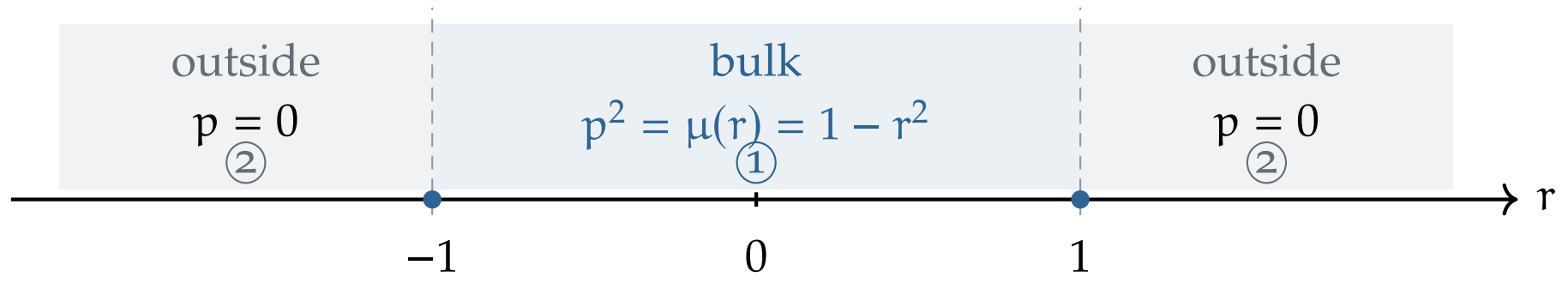

On the diagonal \(\mathbf{r}_1 = \mathbf{r}_2 =\mathbf{r} \), the Fourier phase vanishes. Expanding
the phase-space symbol at fixed $\tau$ gives
\begin{equation}
K_{\text{pert}}(\vect{r},\vect{r} ; \tau)
\simeq
\int \frac{\dd^d {p}}{(2\pi\epsilon)^d}
e^{-\tau\,H(r,p)}
\left[ \mathbb{A}_0(\vect{r},\vect{p}; \tau)
+ \epsilon \mathbb{A}_1(\vect{r},\vect{p}; \tau)
+\cdots\right],
\end{equation}
where $H$ is 
\begin{equation}
H(r,p)=\left(p^2 - \mu(r)\right)^2+ \abs{\sigma(r)}^2,
\end{equation}
 $r=|\vect{r}|$, $p=|\vect{p}|$, and the
coefficients $\mathbb{A}_n$ are matrix-valued. The leading coefficient is normalized as $\mathbb{A}_0=\Id_2$, with $\mathbb{A}_n(\vect r,\vect p;0)=0$ for $n\geq1$.
The $\tau$ integral is dominated by late times, allowing a saddle-point approximation to solve the $p$ integral:
\begin{equation}
  \partial_pH(r,p)=0.
\end{equation}
This leads to two conditions:
\begin{align}
  &\underbrace{p^2 - \mu(r)=0}_{\text{solution \textcircled{1}}}
  &&\text{or} &\underbrace{p=0}_{\text{solution \textcircled{2}}}.
\end{align}
Note that condition \textcircled{1} is a sphere of momentum minima and not an isolated saddle.
On the two solutions, we have
\begin{align}
  &\underbrace{\left.H\right|_{p^2 - \mu(r) =0} = \abs{\sigma}^2}_{\text{solution \textcircled{1}}},
  &
  &\underbrace{\left.H\right|_{p=0} = \mu(r)^2 + \abs{\sigma}^2}_{\text{solution \textcircled{2}}}
    .
\end{align}
Inside the droplet we have $\mu(r) > 0$ and solution \textcircled{1} is dominant because
\begin{equation}
  \left.H\right|_{p=0}>\left.H\right|_{p^2 - \mu(r)=0}.
\end{equation}
Outside the droplet, \(\mu(r)<0\), so $p^2 = \mu(r)$ has no real solution, and solution \textcircled{2} is dominant (see  Figure~\ref{fig:inside-outside}).

The bulk approximations break down at the droplet edge and need to be complemented by a separate edge expansion. In the following, we give the set-up for constructing such an edge expansion, while leaving the actual calculation for future work.  

First, we define $x=R_{\text{cl}}-r$, so that
$\mu(r) = \alpha x + \order{x^2}$, with $\alpha= - \mu'(R_{\text{cl}})>0$.\footnote{For example, for the harmonic potential, $R_{\text{cl}}=1$ and $\alpha=2$.}
Balancing $\epsilon^2\partial_x^2$ with $\alpha x$ gives the Airy length scale \(\ell\):
\begin{align}
	\ell&=(\epsilon^2/\alpha)^{1/3}, &
	\delta &=\alpha\ell, & 
	X&=x/\ell.
\end{align}
For a real gap, the full (i.e. not expanded) \ac{bdg} operator is
\begin{align}
B_\epsilon &= \begin{pmatrix}-h_\epsilon&\sigma_\epsilon\\
\sigma_\epsilon&h_\epsilon\end{pmatrix},
& h_\epsilon &= -\epsilon^2\nabla^2 - \mu(r).
\end{align}
We assume the edge is locally flat at leading order. Fourier transforming the two tangential directions and setting
$\vect q=\vect k_\parallel/\sqrt\delta$ yields
\begin{align}
	\label{eq:simple-edge}
B_\epsilon &\simeq\delta B_{\vect q}, &
B_{\vect q}&=\begin{pmatrix}\partial_X^2-|\vect q|^2+X&S(X)\\
S(X)&-\partial_X^2+|\vect q|^2-X\end{pmatrix},
\end{align}
where, at leading order, $\sigma_\epsilon(R_{\text{cl}}-\ell X) \sim S(X) \delta $ and $B_\epsilon^2\simeq {B_{\vect q}}^2 \delta^2$.
A linearly vanishing bulk gap $\sigma\sim\beta x$ requires
$S(X)\sim(\beta/\alpha)X$ toward the interior.\footnote{The gap profile through the layer can be obtained from the gap equation under the relevant scaling.} The edge scaling matrix heat kernel is then
\begin{equation}
(\partial_s+B_{\vect q}^{\,2})\mathcal K_{\text{edge},\vect q}(s;X,X')=0,
 \quad
 \mathcal K_{\text{edge},\vect q}(0;X,X')=\delta(X-X')\Id_2,
 \quad
 s=\tau\delta^2.
\label{eq:simple-edge-heat}
\end{equation}
It matches the interior and exterior solutions as $X\to\pm\infty$ as follows.
First, choose $a=\ell\Lambda$ in the overlap region, such that
$1\ll\Lambda$ and $a$ are small on the scale of the cloud.
Integrate the bulk density outside the shell $|r-R_{\text{cl}}|<a$,
\begin{equation}
Z_{{\text{bulk}+}\atop\text{outside}}(\tau;a)\simeq 4\pi\left[
\int_0^{R_{\text{cl}}-a}\!\dd r\,r^2 \text{tr}\left[K_{\text{pert}}^{\text{bulk}}(\tau;\vect r,\vect r)\right]
+\int_{R_{\text{cl}}+a}^\infty\!\dd r\,r^2\text{tr}\left[K_{\text{pert}}^{\text{outside}}(\tau;\vect r,\vect r)\right]
\right].
\end{equation}
Here, the superscripts denote the bulk approximation evaluated
inside and outside the cloud, respectively. Their leading large-$\tau$ momentum minima are exactly the solutions \textcircled{1} and \textcircled{2}.
The exterior contribution is kept only to formulate the matching of the heat trace across the turning surface.
Its separate analytic continuation cannot give a physical contribution to the grand potential: in the nonrelativistic theory, the empty vacuum has zero energy. 

The edge kernel lives in the edge region. With the cutoffs chosen above, the region contributes as
\begin{equation}
Z_{\text{edge}}(\tau;\Lambda)\simeq
\frac{4\pi\delta R_{\text{cl}}^2}{(2\pi\epsilon)^2}
\int_{\mathbb R^2}\dd^2q\int_{-\Lambda}^{\Lambda}\dd X\,
\text{tr}\left[\mathcal K_{\text{edge},\vect q}(\tau\delta^2;X,X)\right].
\label{eq:simple-shell}
\end{equation}
The matched perturbative heat trace is their sum.
The cutoff-dependent pieces cancel only after both the bulk and edge expansions have been calculated and matched at the same order.
The resulting computational prescription for the zeta function is then
\begin{equation}
\zeta_{B_\epsilon^2,\text{pert}}(\nu)
\simeq\frac{1}{\Gamma(\nu)}\int_0^\infty\dd\tau\,\tau^{\nu-1}
\left[Z_{{\text{bulk}+}\atop\text{outside}}(\tau;\ell\Lambda)
+Z_{\text{edge}}(\tau;\Lambda)\right]_{\text{matched}}.
\label{eq:simple-zeta}
\end{equation}
The bracket denotes cancellation of cutoff terms at the chosen order in $\epsilon$. 
The edge term is not the divergent part of the bulk integral. It is the finite remainder of the fully renormalized boundary-layer problem after subtracting the appropriate bulk asymptotics.

The integral is defined first where it converges and then analytically continued as in~\cite{Hellerman:2023myh}.
This gives a prescription for computing the perturbative contribution, including the leading edge correction. Higher-order edge corrections require extending the boundary-layer expansion to higher orders.
With this prescription, the perturbative series can be computed in principle to arbitrary order (see Section~\ref{sec:Examples}).

\begin{figure}
  \centering
  \begin{tikzpicture}[x=2.5cm,y=1cm,font=\small]
    \definecolor{bulkblue}{RGB}{45,100,150}
    \definecolor{outergray}{RGB}{100,110,120}
    \definecolor{edgeorange}{RGB}{190,115,35}

    \fill[outergray!9] (-2.15,0.08) rectangle (-1.2,1.5);
    \fill[edgeorange!20] (-1.2,0.08) rectangle (-0.8,1.5);
    \fill[bulkblue!10] (-0.8,0.08) rectangle (0.8,1.5);
    \fill[edgeorange!20] (0.8,0.08) rectangle (1.2,1.5);
    \fill[outergray!9] (1.2,0.08) rectangle (2.15,1.5);

    \node[text=outergray] at (-1.675,1.2) {outside};
    \node[text=bulkblue] at (0,1.2) {bulk};
    \node[text=outergray] at (1.675,1.2) {outside};

    \node at (-1.675,0.78) {$p=0$};
    \node[text=bulkblue] at (0,0.78) {$p^2=1-r^2$};
    \node at (1.675,0.78) {$p=0$};

    \node[text=outergray] at (-1.675,0.35) {\textcircled{2}};
    \node[text=bulkblue] at (0,0.35) {\textcircled{1}};
    \node[text=outergray] at (1.675,0.35) {\textcircled{2}};

    \foreach \edge in {-1,1}{
      \node[text=edgeorange] at (\edge,1.2) {edge};
      \draw[densely dashed,edgeorange!80]
      (\edge,-0.12) -- (\edge,0.55);
    }

    \draw[->,thick] (-2.3,0) -- (2.35,0)
    node[right] {$z$};
    \foreach \pos/\lab in {-1/-1,0/0,1/1}{
      \draw[thick] (\pos,0.055) -- (\pos,-0.055);
      \node[below=5pt] at (\pos,0) {$\lab$};
    }
    \fill[edgeorange] (-1,0) circle (2pt);
    \fill[edgeorange] (1,0) circle (2pt);

    \foreach \edge in {-1,1}{
      \draw[<->,edgeorange,thick]
      ({\edge-0.2},-0.68) -- ({\edge+0.2},-0.68);
      \node[below,text=edgeorange,font=\footnotesize]
      at (\edge,-0.68) {$O(\ell)$};
    }
    \node[text=outergray,font=\footnotesize]
    at (0,-0.68) {$r=|z|$};
  \end{tikzpicture}
  \caption{Bulk and edge regions for the harmonic cloud.
    The interior and exterior momentum minima are marked
    \textcircled{1} and \textcircled{2}, respectively.
    The edge layers near $r=1$ require the
    matrix edge expansion. Their characteristic thickness
    is measured in units of $\ell=(\epsilon^2/2)^{1/3}$.}
  \label{fig:bulk-edge-outside}
\end{figure}
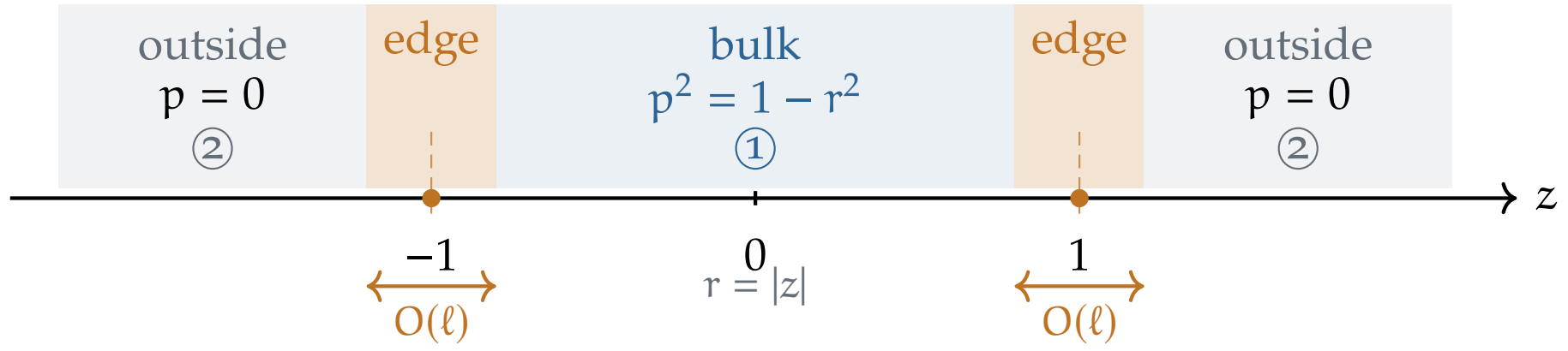

\subsection{Non-Perturbative analysis}
\label{subsec:non-perturbative}

\subsubsection{Hamiltonian mechanics}

We now return to the other rescaling, \(t = \epsilon \tau\), to compute the non-trivial classical trajectories.
Substituting the ansatz (\ref{eq:WKB-ansatz}) into the heat kernel equation and retaining the leading terms in \(\epsilon\), we obtain the Euclidean eikonal equation for the phase $\phi$ of a non-trivial trajectory $\gamma$,
\begin{equation}
\label{eq:eikonal_equation_WKB}
-\partial_t\phi(\vect{r}_1,\vect{r}_2;t)
+\pqty*{- \abs{\nabla\phi(\vect{r}_1,\vect{r}_2;t)}^2 - 1 + V(\vect{r}) }^2
+\sigma(\vect{r})^2=0.
\end{equation}
Here, we picked $\sigma(\vect{r})$ to be real to simplify the notation.
The physical momentum is obtained only after an analytic continuation.
Let
\begin{align}
  \label{eq:euclidean-continuation}
  p &= i\nabla\phi, & s &= -i t, & \phi(\vect{r}_1,\vect{r}_2;t) &= -i \eval*{S(\vect{r}_1,\vect{r}_2;s)}_{s=-i t}.
\end{align}
Then~\eqref{eq:eikonal_equation_WKB} becomes the usual Hamilton--Jacobi equation~\cite{Houchmandzadeh_2020},
\begin{equation}
  \partial_s S(\vect{r}_1,\vect{r}_2;s)+H(|\vect{r}|,p)=0,
\end{equation}
where $H$ now plays the role of a Hamiltonian and is given by
\begin{equation}
  \label{eq:definition-hamiltonian}
  H(\vect{r},p) =\left(p^2-1+V(\vect{r})\right)^2+\sigma(\vect{r})^2.
\end{equation}
We have mapped our problem to standard Hamiltonian mechanics.
The characteristics of the Hamilton--Jacobi equation are classical trajectories in complex phase space, and their actions identify candidate exponentially small contributions.
For a path from \(q(0) = \mathbf{r}_1\) to \(q(s_f) = \mathbf{r}_2\), the action is
\begin{equation}
  \label{eq:definition-WKB-exponential}
  S(\vect{r}_1,\vect{r}_2;s_f)=
  \int_0^{s_f} \dd{s}
  \left[p(s)\cdot\dot q(s)-H(q(s),p(s))\right].
\end{equation}
The Hamiltonian is independent of \(s\), so the energy is conserved,
\begin{equation}
  \pqty*{p^2 + V(r) - 1}^2 + \sigma(r)^2 = E^2 ,
\end{equation}
where $E$ is known as Bogoliubov's quasiparticle energy.
If the potential \(V(\mathbf{r})\) is rotationally invariant (as we assume here), the angular momentum \(L = \abs{ q \times p}\) is conserved and gives another integral of motion.
Isotropy also confines the trajectory to a plane, which is conveniently described in polar coordinates (see Figure~\ref{fig:plane-of-motion}):
\begin{equation}
   q(s) = \pqty*{r(s),\theta(s)}.
\end{equation}
The same decomposition for the momentum gives
\begin{equation}
  \abs*{p(s)}^2 = p_r(s)^2 + \frac{L^2}{r(s)^2} \, .  
\end{equation}
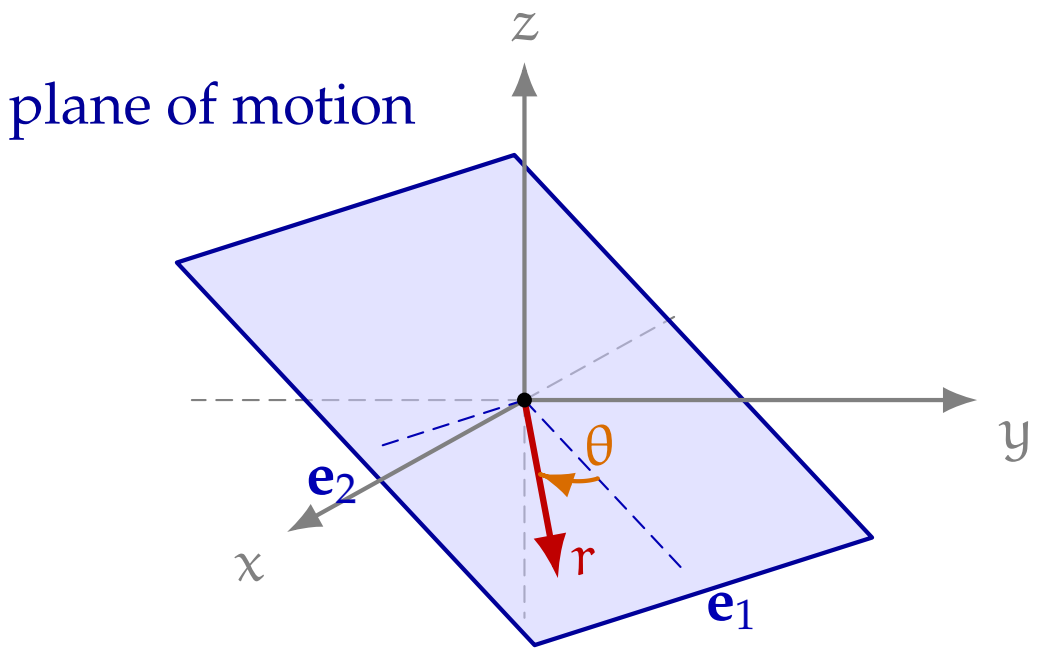
\begin{figure}
\centering
\begin{tikzpicture}[
    x={(-0.45cm,-0.25cm)}, %
    y={(1cm,0cm)},         %
    z={(0cm,0.8cm)},         %
    >=Latex,
    line cap=round,
    line join=round
]

\coordinate (O) at (0,0,0);

\coordinate (A) at (-2.342,-1.121, 1.294);
\coordinate (B) at (-1.822, 1.479,-1.706);
\coordinate (C) at ( 2.342, 1.121,-1.294);
\coordinate (D) at ( 1.822,-1.479, 1.706);

\coordinate (R) at (1.474,0.886,-1.023);

\draw[densely dashed,gray] (O) -- (-2.2,0,0);
\draw[densely dashed,gray] (O) -- (0,-2.2,0);
\draw[densely dashed,gray] (O) -- (0,0,-1.8);

\filldraw[
    fill=blue!20,
    fill opacity=0.55,
    draw=blue!60!black,
    thick
]
(A) -- (B) -- (C) -- (D) -- cycle;

\node[
    blue!60!black,
    align=center
]
at (-0.3,-2.2,2.3)
{plane of motion};

\draw[gray,->,thick]
    (O) -- (3.5,0,0)
    node[below left] {$x$};

\draw[gray,->,thick]
    (O) -- (0,3,0)
    node[below right] {$y$};

\draw[gray,->,thick]
    (O) -- (0,0,2.8)
    node[above] {$z$};

\draw[
    densely dashed,
    blue!70!black
]
(O) -- ({0.234},{1.170},{-1.350})
node[pos=1,below right] {$\mathbf e_1$};
\draw[
    densely dashed,
    blue!70!black
]
(O) -- ({1.2*1.488},{-1.2*0.128},{1.2*0.147})
node[pos=1,below left] {$\mathbf e_2$};

\draw[
    ->,
    very thick,
    red!75!black
]
(O) -- (R)
node[pos=0.70,below right] {$r$};

\draw[
    ->,
    thick,
    orange!85!black
]
plot[
    domain=0:40,
    samples=40,
    variable=\t
]
(
    {0.82*(0.130*cos(\t) + 0.992*sin(\t))},
    {0.82*(0.650*cos(\t) - 0.085*sin(\t))},
    {0.82*(-0.750*cos(\t) + 0.098*sin(\t))}
);

\node[orange!85!black]
at
(
    {0.64*(0.130*cos(-30) + 0.992*sin(-30))},
    {0.64*(0.650*cos(-30) - 0.085*sin(-30))},
    {0.64*(-0.750*cos(-30) + 0.098*sin(-30))}
)
{$\theta$};

\fill (O) circle[radius=1.4pt];
\end{tikzpicture}
\caption{Visualization of the plane of motion spanned by the two basis vectors {\color{blue!70!black}$\mathbf e_1,\mathbf e_2$} shown in three dimensional space {\color{gray}$(x,y,z)\in\mathbb{R}^3$}. We parameterize points in this plane of motion by the coordinates {\color{red!75!black}$r$}, {\color{orange!85!black}$\theta$}.}
\label{fig:plane-of-motion}
\end{figure}
The energy becomes
\begin{equation}
  E^2 = \left(p_r(s)^2+\frac{L^2}{r(s)^2} - \mu(r(s))\right)^2
  +\sigma(r(s))^2,
\end{equation}
which we can solve for $p_r$:
\begin{equation}
  \label{eq:momentum-sheet}
  p_r(s) = \pm p_{r,\pm}(s) =  \pm\sqrt{\pm\sqrt{E^2-\sigma(r(s))^2}
    -\frac{L^2}{r(s)^2} + \mu(r(s))} \, .
\end{equation}
The two signs in the expression of \(p_r(s)\) have different meanings:
\begin{itemize}
\item The inner sign distinguishes particle-like and hole-like branches.
  To see that, consider the limit of vanishing gap and $L=0$, where
  \begin{equation}
    p_r^2 - \mu(r) = \pm E .
  \end{equation}
  The positive sign describes a particle above the local chemical potential; the negative sign describes a hole associated to an unoccupied state below it.
  For nonzero gap, pairing mixes these characteristics, giving particle-like and hole-like quasiparticles.
\item The outer sign determines the sign of the radial momentum. 
\end{itemize}
To determine the direction of propagation, consider the radial
group velocity associated with the positive Bogoliubov energy $E$:
\begin{equation}
  v
  =\frac{\partial E}{\partial p_r}
  =\frac{2p_r}{ E }
  \left(p_r^2+\frac{L^2}{r^2} -\mu(r)\right)
  = \pm\frac{2p_r}{E}
  \sqrt{E^2-\sigma(r)^2},
\end{equation}
where the last sign is the inner sign of (\ref{eq:momentum-sheet}).
For real propagating modes, the velocity  has the same sign as the momentum on the particle-like branch and the opposite sign on the hole-like branch.
Switching between these branches can thus reverse the motion without reversing the momentum, as in the Andreev reflection~\cite{Andreev1964}.%
\footnote{For the complex zero-energy trajectories, the particle--hole labels are understood by analytic continuation.}

\subsubsection{Trajectories}

What are the relevant trajectories for our problem?
We want to compute the Mellin integral in Eq.~\eqref{eq:B2-zeta}: 
\begin{equation}
  \zeta_{B^2}(s)
  =
  \frac{\epsilon^{-s}}{\Gamma(s)}
  \int_0^\infty \frac{\dd{t}}{t}\, t^s
  \int \dd^3{r}\,
  \Tr  K_{B^2}(\vect{r},\vect{r};t)
\end{equation}
with the ansatz in Eq.~\eqref{eq:WKB-ansatz}:
\begin{equation}
  K_{B^2}(\vect{r}_1,\vect{r}_2;t)
  =
  \sum_\gamma
  e^{-\phi_\gamma(\vect{r}_1,\vect{r}_2;t)/\epsilon}   \mathbb{P}_{\gamma,\epsilon}(\vect{r}_1,\vect{r}_2;t) .
\end{equation}
Both the time and space integrals can be solved by saddle point, which translates into conditions on the leading trajectories.
\begin{itemize}
\item The integrand is evaluated at \(\vect{r}_1=\vect{r}_2=\vect{r}\), so we need trajectories \(\vect{q}(s)\) that are closed in configuration space.
\begin{equation}
  \vect{q}(0)=\vect{q}(s_f).
\end{equation}
Equivalently,
\begin{align}
	r_1&=r_2=r,&
  \theta_1&=\theta_2+2n\pi,
&
  n&\in\setZ.
\end{align}
Using
\begin{equation}
  p\cdot \dd{q} = p_r \dd{r} + L \dd{\theta}
\end{equation}
the action is written as
\begin{equation}
  S
  =
  \int_{r_1}^{r_2}\dd{r} p_r
  +L\int_{\theta_1}^{\theta_2} \dd{\theta}
  - E^2 \cdot s_f.
\end{equation}
So, for our trajectories \(\gamma\), after analytic continuation
\begin{equation}
  \phi_\gamma (t)
  =
  E^2 \cdot t
  -i\left[
    \oint_\gamma p_r \dd{r} + 2\pi nL
  \right].
\end{equation}
The trajectory \(\gamma\) is closed, so if we vary \(t\), \(E\) and \(L\) must also change.
In this sense they are functions of \(t\).
Using Hamilton's equations
\begin{align}
  \frac{\dd r}{\dd s}
  & =4\underbrace{\left(p_r^2+\frac{L^2}{r(s)^2} - \mu(r(s))\right)}_{ \pm \sqrt{E^2 - \sigma(r(s))^2}}p_r, &\frac{\dd \theta}{\dd s}
&=\frac{\partial H}{\partial L}
= 4 \left(\pm\sqrt{E^2 - \sigma(r(s))^2}\right)\frac{L}{r^2},
\end{align}
we obtain an implicit form for the functions \(E(t)\) and \(L(t)\):
\begin{align}
    s_f=\int_0^{s_f}\dd{s}
& =\int_{r_1}^{r_2}\dd{r} 
\frac{1}{4\left(\pm\sqrt{E^2 - \sigma(r)^2}\right)p_r},
& \theta_2-\theta_1 &=\int_{\theta_1}^{\theta_2}\dd{\theta}
=\int_{r_1}^{r_2}\dd{r}\frac{L}{r^2p_r}.
\end{align}
\item The integral over \(\vect{r}\) is solved by saddle point, so we need to identify the trajectories \(\gamma\) that satisfy \(\delta\phi=0\) for a variation \(\delta\vect{r}\) at fixed \(t\).
For a generic classical trajectory, from \(\vect{r}_1\) to \(\vect{r}_2\), the variation can only be a boundary term:
\begin{equation}
  \delta\phi
  =
  -i\vect{p}_2\cdot\delta\vect{r}_2
  +i\vect{p}_1\cdot\delta\vect{r}_1.
\end{equation}
Since for us \(\vect{r}_1=\vect{r}_2=\vect{r}\), we conclude that
\begin{equation}
  \vect{p}_2=\vect{p}_1.
\end{equation}
We see that the leading trajectories are periodic in phase space, not just in configuration space.
\item The final integral is over \(t\). Again, by saddle point, we have the condition
\begin{equation}
  \frac{\dd \phi_\gamma}{\dd t}=0.
\end{equation}
However, for a periodic stationary trajectory, this is the equation that defines \(E(t)\):
\begin{equation}
  \frac{\dd \phi_\gamma}{\dd t} = E(t)^2 = 0.
\end{equation}
\end{itemize}
We conclude that the saddles of the Mellin integral are periodic, zero-energy trajectories
that correspond to nonlocal contributions to the energy.%
\footnote{The local perturbative sector is associated with the short-time branch near coincident endpoints.} 
The zero-energy condition selected by the proper-time analysis has a quasiparticle interpretation.
Recall that $E$ is Bogoliubov's quasiparticle energy.
Whenever the gap is nonzero, real propagating modes require $ E^2 \geq \sigma(r)^2$, so at $E=0$ the momentum is 
complex in the bulk.
The relevant trajectories describe quasiparticle propagation analytically continued into a classically forbidden region, as in ordinary \ac{wkb} tunneling: they are complex characteristics of the squared \ac{bdg} operator, rather than real-time trajectories of zero-energy particles.

\subsubsection{Zero-energy \(L=0\) orbits in a spherically symmetric potential}

We now specialize to the radial motion in a spherically symmetric potential $V(r)=r^p$, with $p$ even.
Cycles with \(L = 0\) are stationary under variations of the angular momentum.
Using Eq.~\eqref{eq:momentum-sheet}:
\begin{equation}
    \odv{p_r}{L} = - \frac{L}{p_r r^2}
\end{equation}
so the variation of the radial action is
\begin{equation}
  \odv{}{L} \oint_{\gamma} p_r \dd{r} = - \oint \frac{L}{r^2 p_r} \dd{r} .  
\end{equation}
Because of this, in the following we will focus on \(L = 0\) trajectories.
Stationarity of the cycle does not in general imply that the nearest Borel singularity has zero angular momentum; we will make this assumption and then use numerical evidence to substantiate it.

We use the leading bulk gap profile
\begin{equation}
    \sigma(r) = y_0 \mu(r) .
\end{equation}
Higher-order terms in the \(\epsilon\) expansion are not needed to determine the leading semiclassical action; they affect only the associated perturbative expansion in the transseries, encoded in \(\mathbb{P}_{\gamma,\epsilon}\) in Eq.~\eqref{eq:WKB-ansatz}.

At zero energy, the spectral curve obtained from the eikonal equation factorizes into two components:
\begin{equation}\label{eq:spectral-curve}
  \pqty*{p_r^2 - \mu(r)}^2
  +y_0^2 \mu(r)^2 = 
  \bqty*{p_r^2 - \pqty*{1 + i y_0} \mu(r) }
    \bqty*{p_r^2 - \pqty*{1 - i y_0} \mu(r) }
  =0.
\end{equation}
There are four momentum sheets, \(\pm p_{r,\pm}\), corresponding to the independent choices of the inner branch
and the outer square-root sign
\begin{equation}
   p_{r,\pm}^2 =  \pqty*{1\pm i y_0} \mu(r) \, . 
\end{equation}
These sheets are analytic continuations of the particle-like
and hole-like Bogoliubov branches, each with two possible signs of
the radial momentum. Their physical labels are inherited from
the real propagating regime: particle-like excitations have
velocity parallel to their momentum, whereas hole-like
excitations have velocity antiparallel to it.
An excitation can reverse its motion by changing its
particle--hole character without reversing its momentum.
At complex zero energy, however, these labels specify
analytic branches rather than directions of real propagation.

Each factor of the spectral curve defines a two-sheeted curve with turning points
at $V(r)=1$, namely the $p$ roots of $r^p=1$. At these points,
all four momenta vanish. At the physical droplet edge \(r = \pm 1\), the
leading gap and the local chemical potential vanish together,
allowing the zero-energy curve to reach a turning point.
The remaining complex roots are turning points of the
analytically continued problem, rather than additional
physical boundaries of the cloud.

The intersections of the two components in Eq.~\eqref{eq:spectral-curve} are degenerate.
The coincidence of their momentum values does not
uniquely determine how a trajectory continues between sheets.
To resolve the connections, we introduce a small
nonzero energy, which splits the degenerate edge branch point into distinct branch points. At zero angular momentum, Eq.~\eqref{eq:momentum-sheet} reads
\begin{equation}
    p_r = \pm\sqrt{\mu(r) \pm \sqrt{E^2 - \sigma(r)^2}}.
\end{equation}
This separates two kinds of branch points.
\begin{itemize}
\item When the inner square root vanishes,
\begin{align}
  E^2 &= \sigma(r)^2, & p_r^2&= \mu(r),
\end{align}
the particle-like and hole-like branches meet, generally at
nonzero momentum. Analytic continuation around such a branch
point exchanges these branches. This is the spectral-curve
counterpart of Andreev conversion.
\item When the outer square root vanishes,
\begin{align}
	p_r&=0, & E^2 & = \mu(r)^2 + \sigma(r)^2,   
\end{align}
two opposite-momentum branches meet. Their exchange is the
counterpart of ordinary momentum reflection. 
\end{itemize}
Closed cycles
may involve either type of connection or combinations of both.
We determine the cycles at nonzero energy by
tracking the momentum continuously around the resolved branch
points, including its return to the initial sheet. We then
take the limit $E\to0$ along the resulting family of cycles.
This procedure fixes the sheet connections without assigning
arbitrary sign changes at the degenerate zero-energy
intersections.

For a closed zero-energy cycle, the Euclidean \ac{wkb} phase is
\begin{equation}
    \phi_\gamma^*=-i\oint_\gamma p_r dr .
\end{equation}
It measures the action accumulated along the complex
quasiparticle trajectory.%
\footnote{The zero-energy cycles are understood as limits of cycles on the
  resolved finite-energy curve. For \(E \to 0\) the limiting action is finite even
though the period is singular.}
For the power-law potentials considered here, the required
zero-energy integrals can be evaluated in terms of a beta
function.

Let $r_A$ and $r_B$ be the turning points of a given closed trajectory.
By definition, $r_A^p=r_B^p=1$, and we define the coefficient \(J_p\) from
\begin{equation}
  \label{eq:Beta-integral}
\int_{r_A}^{r_B}\sqrt{1-r^p}\,dr
=\frac{r_B-r_A}{p}
 B\lrp*{\tfrac{1}{p},\tfrac{3}{2}} = \pqty*{r_B - r_A} J_p.
\end{equation}
The action of each cycle follows by combining these integrals
with the appropriate factors $\pm\sqrt{1\pm i y_0}$. Each such action provides a candidate exponential weight
$\exp(-\phi_\gamma^*/\epsilon)$.
Its real part determines
exponential suppression, its imaginary part produces
oscillations, and its modulus is used to estimate the optimal truncation of the asymptotic perturbative expansions.

An important feature of fermionic systems is Andreev reflection. %
Close to the Fermi surface, a particle-like
quasiparticle can bounce back as a hole-like quasiparticle: its group velocity
reverses while its canonical momentum changes only slightly~\cite{Andreev1964}.
Here we consider its analytic continuation.
A cycle with a
particle-like outgoing leg and a hole-like returning leg
involves the difference of their momentum integrals, so the common Fermi-momentum contribution cancels. This explains how cancellation between the two branches
can produce an action smaller than that of a
momentum-reversing cycle along the same spatial contour where momenta add up.
Andreev reflection therefore gives the smallest action among these connections along the same contour, and is the candidate controlling optimal truncation.
For a \(V = r^p\) potential, we can therefore evaluate the candidate minimal action by connecting a pair of nearest turning points:
\begin{multline}\label{equ:OT-action}
  \abs*{\phi_{\gamma}^{*}} = \abs*{\pqty{r_A - r_B} J_p \sqrt{1 + i y_0} + \pqty{r_B - r_A} J_p\sqrt{1 - i y_0}} \\
  = \frac{4}{p} B\lrp*{\tfrac{1}{p},\tfrac{3}{2}}  \sin*(\tfrac{\pi}{p})  \operatorname{Im} \sqrt{1 + i y_0} \, .
\end{multline}
Except for the harmonic potential \(p = 2\), this candidate leading cycle closes in complex configuration space rather than along the physical diameter.
Zero quasiparticle energy is the condition that the worldline sits at the gap edge, i.e. at the boundary between the propagating and evanescent regimes, which is where the Andreev reflection happens.

\subsection{Numerics from the perturbative expansion}\label{Section:perturbative-coeff-numerics}

In the following we present numerical results to test our \ac{wkb} analysis.

\subsubsection{Extraction of the perturbative coefficients}

We now compute the perturbative coefficients whose large-order
behavior will test the actions obtained above. Our starting
point is the bulk heat-kernel expansion of Section~\ref{sec:perturbative}.
Following~\cite{Hellerman:2023myh}, we insert the phase-space
expansion into the heat equation and expand the Moyal products
in powers of $\epsilon$. We then solve recursively for the gap and the heat kernel.

The symbol \(\mathcal{K}_{\text{pert}}\) takes the form %
\begin{equation}\label{equ:perturbative-heat-kernel-integrand}
 \mathcal K_{\text{pert}}(r,p ; \tau)
 =\sum_{j\geq0}\epsilon^j e^{-\tau H(r,p)} \mathbb{A}_j(r,p ; \tau) ,
\end{equation}
where $\mathbb{A}_j$ is a matrix-valued polynomial in $\tau$,
with coefficients depending on position and momentum.
The proper-time integral can be evaluated
term by term in terms of gamma functions.

The gap is determined self-consistently by imposing its
saddle-point equation order by order.
The leading term gives $\sigma(\mathbf r)=y_0\mu(\mathbf r)$, reproducing the \ac{lda}.
Odd-order contributions to the grand potential vanish after momentum integration.

The grand potential takes the form
\begin{equation}
 \frac{\Omega}{N\mu}
 \sim \epsilon^{-3}\sum_{n=0}^{\infty}a_n\epsilon^{2n}.
\end{equation}
After removing the common phase-space factor $\epsilon^{-3}$,
the integrated coefficients are
\begin{equation}
\begin{split}
 a_n={}&-\frac12\operatorname{FP}
 \int d^3r\,
 \biggl[
 \frac{1}{\Gamma(s)}
 \int_0^\infty d\tau\,\tau^{s-1}
 \int\frac{d^3p}{(2\pi)^3}\,
 \operatorname{tr}
 \mathcal K_{n}(\tau,\mathbf r,\mathbf p)
 \biggr]_{s=-1/2} ,
\end{split}
\label{eq:perturbative-coefficient-extraction}
\end{equation}
where $ \mathcal K_{2n}(\tau,\mathbf r,\mathbf p)$ is the $\epsilon^{2n}$ coefficient in \eqref{equ:perturbative-heat-kernel-integrand}. The brackets denote the analytic continuation of the Mellin
transform, with the unitarity subtraction understood.
The spatial finite part, denoted by $\operatorname{FP}$,
is defined by analytic continuation of the radial integrals
at the cloud edge. To analyze their large-order growth, we compute approximately twenty coefficients using this procedure, through order $\epsilon^{40}$.

For the nonterminating bulk expansions considered below, the coefficients exhibit factorial growth, \( a_n \propto (2n)! / \phi_\gamma^{2n} \). At fixed small $\epsilon$,
successive terms initially decrease because of the suppression
by $\epsilon^{2n}$. Eventually, factorial growth dominates and
the terms begin to increase. The optimal truncation is obtained
by retaining terms up to the smallest term. This occurs when
\begin{equation}
 \left|
 \frac{a_{n+1}\epsilon^{2n+2}}
      {a_n\epsilon^{2n}}
 \right|
 \sim \frac{4n^2\epsilon^2}{|\phi_\gamma|^2}\simeq1,
\end{equation}
giving the leading estimate
\begin{equation}\label{equ:optimal-truncation}
 N_{\text{opt}}\simeq\frac{|\phi_\gamma|}{2\epsilon}.
\end{equation}
Here $N_{\text{opt}}$ denotes the coefficient index: the corresponding
term is proportional to $\epsilon^{2N_{\text{opt}}}$. The potential determines the complex trajectories and their actions, and thus changes the predicted truncation order. Using Eq.~\eqref{equ:OT-action}, the nearest Andreev cycle for a potential \(V = r^p\) gives
\begin{equation}\label{equ:optimal-truncation-andreev}
   N_{\text{opt}} = \frac{2}{p \epsilon} B\lrp*{\tfrac{1}{p},\tfrac{3}{2}}  \sin*(\tfrac{\pi}{p})  \Im \sqrt{1 + i y_0} \, .
\end{equation}
The connection with nonperturbative effects becomes apparent
through the Borel transform of the series,
\begin{equation}
 B(s)=
 \sum_{n=0}^{\infty}\frac{a_n}{(2n)!}\,s^{2n}.
\end{equation}
Dividing out the factorial growth produces a series with a
finite radius of convergence, determined by the distance $|\phi_\gamma|$
to the nearest singularities in the Borel plane. Their locations
can be complex: writing one such location as
$\phi_\gamma=|\phi_\gamma|e^{i\theta}$, its modulus controls the growth,
while its phase controls the oscillations of the coefficients.

If the analytic continuation and Laplace integral
exist, Borel resummation defines a function with the original
asymptotic expansion. The associated nonperturbative sectors
carry exponential factors of the form $e^{-\phi_\gamma/\epsilon}$.
The actions governing these sectors also determine the
large-order growth and the optimal truncation of the
perturbative series. We will estimate the optimal truncation
in two ways: from the nonperturbative actions and 
from the growth of the perturbative coefficients.

\subsubsection{Fit of the perturbative coefficients}
We analyze the computed coefficients in two steps, separating
their factorial growth from their sign oscillations. The fitting
ansatz is motivated by the large-order behavior associated
with complex Borel singularities. A singularity at
$\Phi=Ae^{i\theta}$ contributes a term proportional to
$\Gamma(2n+\beta)\Phi^{-2n}$. Combining it with its complex
conjugate produces an oscillation. We therefore use the model 
\begin{equation}
 a_n^{\text{fit}}
 =
 C\,\mathcal E_n(A,\beta,c)\cos(2n\theta-\delta),
 \qquad
 \mathcal E_n(A,\beta,c)
 =
 \frac{\Gamma(2n+\beta)}{A^{2n}}
 \exp\left(\frac{c}{2n+\beta}\right).
 \label{eq:perturbative-fit-ansatz}
\end{equation}
The amplitude $\mathcal E_n$ describes the factorial growth.
Its scale $A$ corresponds to the distance of the singularities
from the Borel origin, while their angle $\theta$ determines
the oscillation. The parameters $\beta,c$ describe corrections
to the growth, and $C,\delta$ determine the amplitude and phase~\cite{Marino:2012zq,Marino:2023resurgence,Aniceto:2018bis}.

We will first determine the growth parameters $A,\beta,c$ from logarithmic
ratios of coefficients. We choose a separation
$j$ between coefficient indices that eliminates the remaining fit parameters:
\begin{equation}
 \log\left|\frac{a_n}{a_{n-j}}\right|.
\end{equation}
We use $j=2$ for the quartic potential and $j=3$ for the sextic potential. These ratios eliminate the overall normalization and cancel the oscillatory magnitude, isolating the growth parameters.

In the second step, we hold the fitted amplitude fixed and
divide the signed coefficients by it:
\begin{equation}
\tilde a_n=\frac{a_n}{\mathcal E_n}
 \simeq C\cos(2n\theta-\delta).
\end{equation}
We then determine $C,\theta,\delta$ from the remaining
oscillation. This procedure gives separate estimates of the singularity modulus and angle, which we compare with the \ac{wkb} predictions.
\section{Examples at $L=0$}\label{sec:Examples}

The preceding section treated a general confining potential. We now consider specific radial monomial potentials, using the leading approximation
\begin{align}
  \sigma(r) &= y_0\mu(r) && \text{with } y_0=1.1622.
\end{align}
We begin with the reference case \(V=0\), where Andreev reflection is absent. We then discuss the harmonic potential \(V(r)=r^2\), which displays an exceptional behavior: cancellations leave only three perturbative terms. Finally, for \(V(r)=r^4\) and \(V(r)=r^6\), we construct the nonperturbative saddles and compare the predicted optimal truncation with numerical estimates.

In both nonterminating examples, the coefficients generated from the microscopic expansion identify the modulus and angular pattern predicted by the Andreev cycles connecting adjacent turning points.
This comparison tests the semiclassical construction without using its actions to generate or fit the coefficients.
It concerns the leading large-order behavior of the specified perturbative bulk observable; the amplitudes of individual nonperturbative sectors require further analysis. 

Note that our Hamilton--Jacobi analysis allows us to determine the cycles of the spectral curve.
However, we must distinguish these cycles from the elementary instantons, which describe the semiclassical configurations of the physical system, and from the Borel singularities that control the large-order behavior of the expansion of a particular physical observable, \emph{i.e.} the grand potential.
We will encounter evidence both for cycles that are not independent instantons (harmonic potential) and cycles that are instantons with zero Stokes coefficient that do not control the asymptotic behavior of the perturbative expansion (sextic potential).
A more detailed analysis goes beyond the scope of this work.

\subsection{$V(r)=0$}
For starters, let us consider a gas confined inside a sphere of radius \(R\) without potential, \(V=0\).
Assuming a constant gap, \(\sigma=y_0\), also the momenta are constant,
\begin{equation}
  p_r=\pm\sqrt{1\pm i y_0 } .
\end{equation}
The simplest periodic trajectory crosses the
interval and returns after momentum reversal. Its total length is $4R$.
Here the reversal is imposed by the hard walls, rather than a zero of
the gap and there is no Andreev reflection.
The corresponding action is
\begin{align}
 \phi_\gamma^* &= \pm4Ri\sqrt{1\pm i y_0}, & \abs{\phi_\gamma^*} = 4R \pqty*{1+y_0^2}^{1/4}
 =4.9529 R.
\end{align}
In Appendix~\ref{sec:CubeBall} we recover the same result via a detailed analysis of the ball with Dirichlet \ac{bc}.

\subsection{$V(r)=r^2$}
\label{sec:vr=r2}
Next we consider a harmonic trap, \(V(r)=r^2\).

\subsubsection{Zero-energy trajectories}

\begin{figure}
  \centering
  \includegraphics[width=\linewidth]{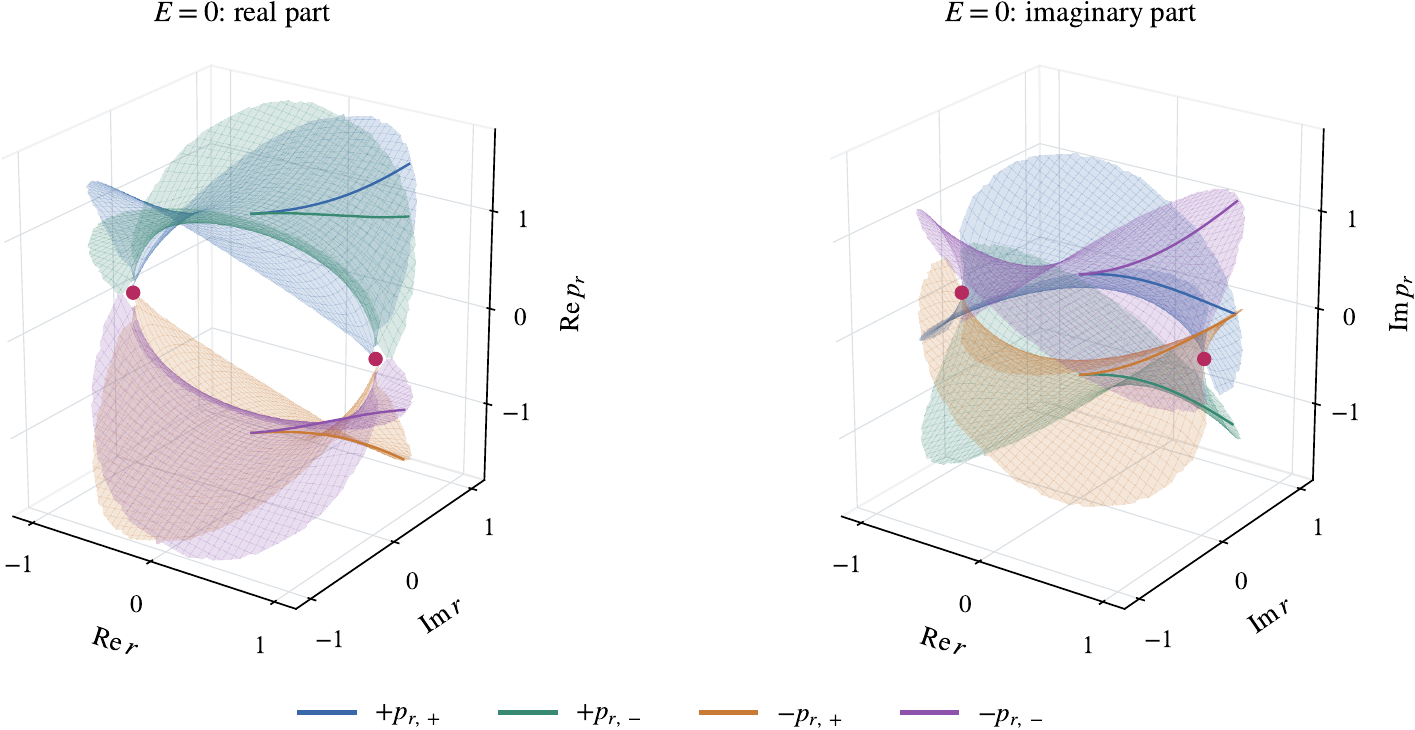}
  \par\vspace{3mm}
  \includegraphics[width=\linewidth]{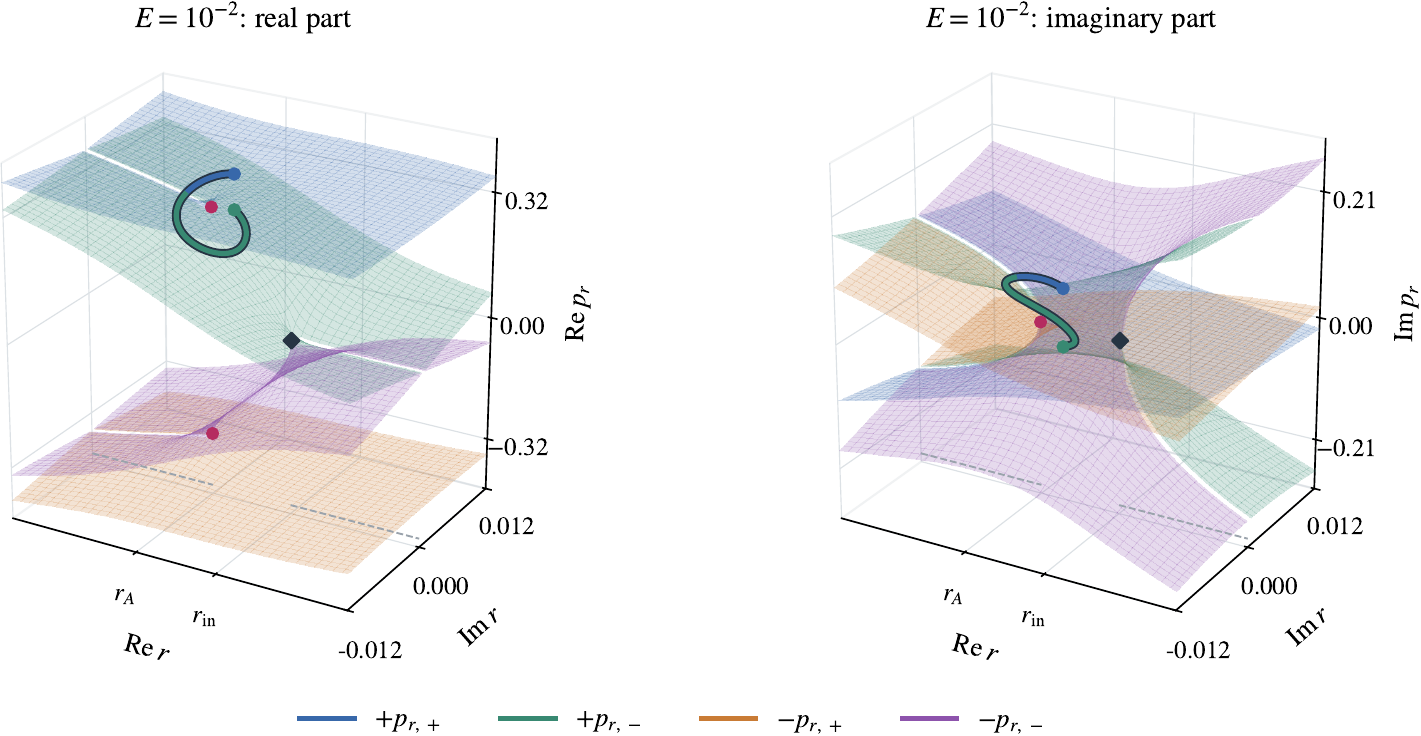}
  \caption{Harmonic momentum sheets over the complex position plane,
    showing $\text{Re } p_r$ on the left and $\text{Im } p_r$ on the right.
    Top: the zero-energy surfaces, with the junctions
    $(r,p_r)=(\pm1,0)$ marked in red.
    Bottom: the resolved neighborhood of the positive edge at
    $E=10^{-2}$. Red points mark the inner-root junction at $p_r\ne0$;
    the black diamond marks the ordinary turning point at $p_r=0$.
    The blue-to-green lifted path connects $+p_{r,+}$ to $+p_{r,-}$,
    reversing the sign of the imaginary momentum while its real part
    remains positive. It is a local sheet connection, rather than a
    complete closed orbit. Crossings in a single projection need not
    be coincidences of the full complex momentum.}
  \label{fig:sec5-harmonic-sheets}
\end{figure}

Here, the sheet junctions are at \(r=\pm 1\), so there is only one endpoint pair, and the closed trajectories describe twice the diameter of the droplet in position space.
This structure is particularly simple, and we can discuss the closed zero-energy trajectories in detail.
The action of a diameter cycle is
\begin{equation}
  \phi_\gamma^*
  = -i\int_{-1}^{1} \dd{r} \left(p_{r, \rightarrow}-p_{r, \leftarrow}\right) 
  = -2 i J_2\left[
      \left(s_{\rightarrow}\sqrt{1+i\eta_{\rightarrow}y_0}\right)
      -\left(s_{\leftarrow}\sqrt{1+i\eta_{\leftarrow}y_0}\right)
    \right],
\end{equation}
where \(J_2 = \pi/4\) is defined in Eq.~(\ref{eq:Beta-integral}), and \(s_{\substack{\leftarrow\\[-2pt]\rightarrow}}\) and \(\eta_{\substack{\leftarrow\\[-2pt]\rightarrow}}\) are signs.
We start with \(s_{\rightarrow}=\eta_{\rightarrow}=+1\), \emph{i.e.} a particle moving in the positive direction.
At the edge, there are three possible sheet changes (see Figure~\ref{fig:sec5-harmonic-sheets}):
\begin{enumerate}
\item Andreev reflection \(+p_{r,+} \to + p_{r,-}\). The corresponding action is
  \begin{equation}
    \left.\phi_\gamma^*\right|_{\text{Andreev}}
    = -i\frac{\pi}{2}
    \left(\sqrt{1+i y_0}-\sqrt{1-i y_0}\right)
    = \pi \Im \sqrt{1 + i y_0}.
  \end{equation}
  Figure~\ref{fig:sec5-harmonic-sheets} illustrates the 
  momentum surfaces and their finite-energy resolution.
  The local blue-to-green continuation exchanges $+p_{r,+}$ and $+p_{r,-}$ at
  nonzero momentum: $\Im p_r$ changes sign while $\Re p_r$ remains
  positive.
  A second endpoint connection is required to close the diameter cycle.
\item Ordinary reflection \(+p_{r,+} \to - p_{r,+}\) with action
  \begin{equation}
    \left.\phi_\gamma^*\right|_{\text{refl.}}
    = -i\frac{\pi}{2}
    \left(\sqrt{1+i y_0}-\left(-\sqrt{1+i y_0}\right)\right) 
    = -i\pi\sqrt{1+i y_0}.
  \end{equation}
\item Combined Andreev and ordinary reflection \(+p_{r,+}\to-p_{r,-}\), with action
  \begin{equation}
    \left.\phi_\gamma^*\right|_{\text{mix}}
    = -i\frac{\pi}{2}
    \left(\sqrt{1+i y_0}-\left(-\sqrt{1-i y_0}\right)\right) 
    = -i\pi \Re\sqrt{1+i y_0}.
  \end{equation}
\end{enumerate}
The names above are, however, just labels: we are discussing complex trajectories, not actual movements in position space. In particular, the combined reflection 3, \((+p_{r, +}\to-p_{r, -})\), has no interpretation as a real classical trajectory.
In fact, while the three cycles are geometrically distinct, their
zero-energy \ac{wkb} actions are not independent.
Denoting the Andreev,
ordinary-reflection, and combined connections by $A$, $B$, and $C$,
respectively, their actions satisfy
\begin{equation}
  \phi_{\text{combined}}^* = \phi_{\text{reflection}}^* - \phi_{\text{Andreev}}^*.
\end{equation}
This suggests that the combined connection could be associated with a relative action between the Andreev and ordinary-reflection sectors.
This would mean that it does not represent an independent instanton and need not produce a Borel singularity of the perturbative expansion.

\begin{figure}[t]
  \centering

  \begin{tikzpicture}[scale=1,font=\footnotesize]

    \def\rinner{1.6221}
    \def\rmiddle{3.5356}
    \pgfmathsetmacro{\router}{
      sqrt(\rinner^2+\rmiddle^2)
    }
    \def\lim{4.25}

    \draw[step=2,gray!9,thin]
    (-\lim,-\lim) grid (\lim,\lim);

    \draw[gray!45,thin]
    (-\lim,0) -- (\lim,0);
    \draw[gray!45,thin]
    (0,-\lim) -- (0,\lim);

    \foreach \r in {\rinner,\rmiddle,\router} {
      \draw[gray!45,thin,densely dashed]
      (0,0) circle[radius=\r];
    }

    \foreach \x in {-\rinner,\rinner} {
      \fill[pointgreen]
      (\x,0) circle[radius=2.8pt];
    }

    \foreach \y in {-\rmiddle,\rmiddle} {
      \fill[pointgray]
      (0,\y) circle[radius=2.2pt];
    }

    \foreach \x in {-\rinner,\rinner} {
      \foreach \y in {-\rmiddle,\rmiddle} {
        \fill[pointgray]
        (\x,\y) circle[radius=2.2pt];
      }
    }

    \foreach \x in {-4,-2,0,2,4} {
      \draw[gray]
      (\x,-\lim) -- ++(0,-0.07);
      \node[below,font=\small]
      at (\x,-\lim-0.07) {$\x$};
    }

    \foreach \y in {-4,-2,0,2,4} {
      \draw[gray]
      (-\lim,\y) -- ++(-0.07,0);
      \node[left,font=\small]
      at (-\lim-0.07,\y) {$\y$};
    }

    \draw[gray,thin]
    (-\lim,-\lim) rectangle (\lim,\lim);

    \node[below=16pt] at (0,-\lim)
    {$\Re \phi_\gamma^*$};

    \node[rotate=90,above=19pt] at (-\lim,0)
    {$\Im \phi_\gamma^*$};
    \circlearrows[0.35]{0}{4.03}{180}{0}{mPm}{pPp}
    \circlearrows[0.35]{0}{3.03}{180}{0}{mPp}{pPm}

    \circlearrows[0.35]{1.62}{0.5}{180}{0}{pPp}{pPm}
    \circlearrows[0.35]{1.62}{-0.5}{180}{0}{mPm}{mPp}

    \circlearrows[0.35]{-1.62}{0.5}{180}{0}{mPp}{mPm}
    \circlearrows[0.35]{-1.62}{-0.5}{180}{0}{pPm}{pPp}
    
    \circlearrows[0.35]{2.12}{4.03}{180}{0}{mPm}{pPm}
    \circlearrows[0.35]{-2.12}{4.03}{180}{0}{mPp}{pPp}
    \circlearrows[0.35]{2.12}{-4.03}{180}{0}{pPp}{mPp}
    \circlearrows[0.35]{-2.12}{-4.03}{180}{0}{pPm}{mPm}

    \circlearrows[0.35]{0}{-4.03}{180}{0}{pPm}{mPp}
    \circlearrows[0.35]{0}{-3.03}{180}{0}{pPp}{mPm}

    \begin{scope}[shift={(5.0,2.0)}]

      \foreach \col/\lab/\y in {
        pPp/{+p_{r,+}}/0,
        pPm/{+p_{r,-}}/-0.65,
        mPp/{-p_{r,+}}/-1.30,
        mPm/{-p_{r,-}}/-1.95
      } {
        \draw[\col,line width=2.5pt]
        (0,\y) -- (0.8,\y);

        \node[right]
        at (1.0,\y) {$\lab$};
      }

    \end{scope}
  \end{tikzpicture}
  \caption{The eight distinct nonzero harmonic \ac{wkb} action values.
    Green circles show the real pair of smallest modulus, obtained by
    inner-sheet exchange along the diameter (Andreev reflection). Gray points correspond to
    the other sheet connections along the same spatial contour.
    Closed paths are identified by the sheets in the two directions \(-1 \to 1 \) and \(1 \to -1\).
    Each point can describe more than one process.}
  \label{fig:sec5-harmonic-actions}
\end{figure}
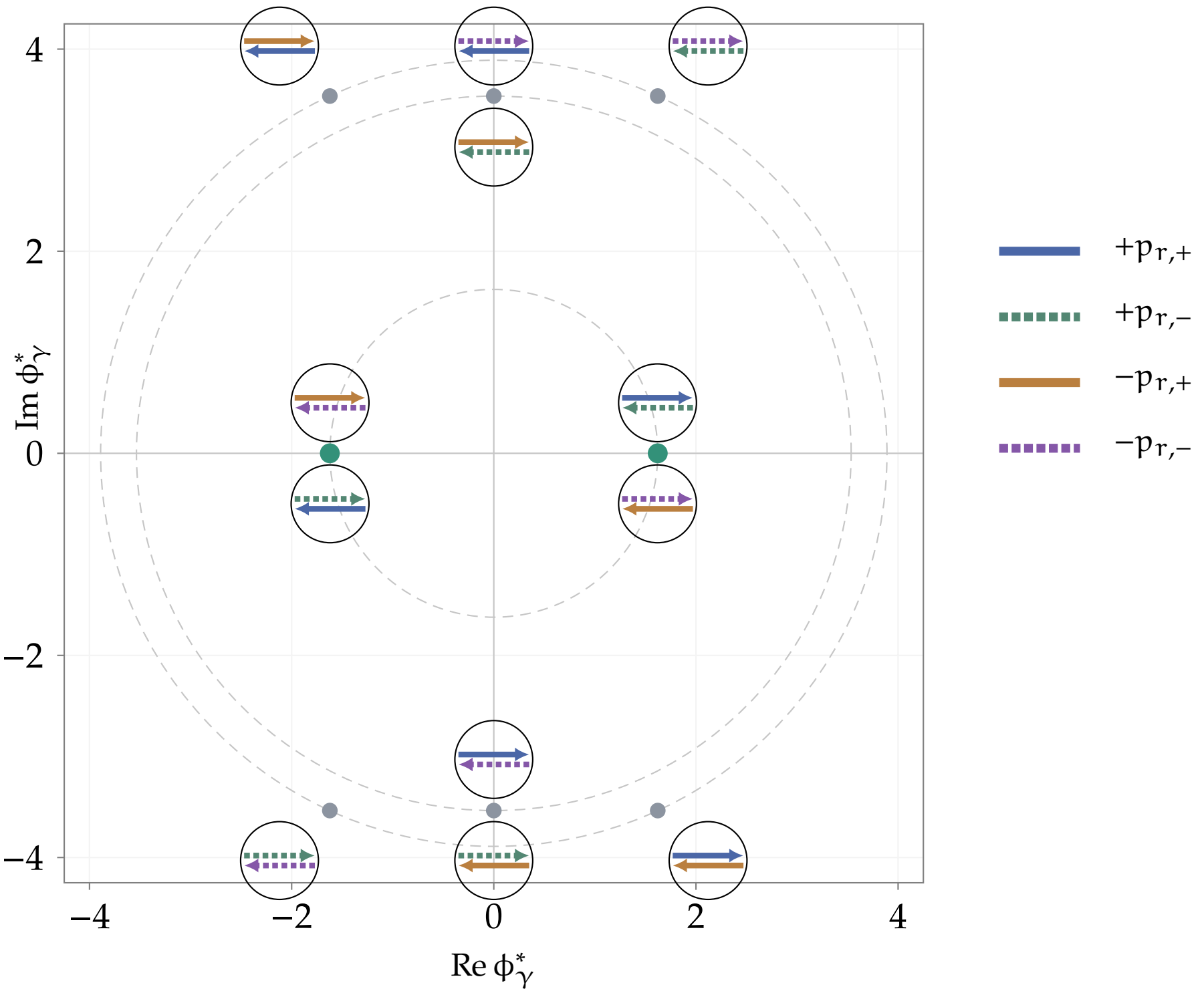

The possible trajectory combinations lead to eight values of the stationary action (see Figure~\ref{fig:sec5-harmonic-actions}).
Since \(\abs{\Im \sqrt{1 + i y_0}} < \abs{\Re \sqrt{1 + i y_0}} < \abs{\sqrt{1 + i y_0}} \), those with the smallest modulus correspond to Andreev reflection.
Plugging in \(y_0 = 1.1622\), the numerical values of the action are \(\pm 1.6221\) (Andreev), \(\pm1.6221 \pm 3.5356 i \) (ordinary), and \(\pm3.5356 i\) (combined).

\subsubsection{Perturbative expansion}

As discussed in Section~\ref{Section:perturbative-coeff-numerics}, we can compute the perturbative expansion to any given order.
For the harmonic potential, the result is
\begin{equation}
  \frac{\Omega}{N\mu}
  = a_0 \pqty*{\frac{2\mu}{ \omega}}^{3}
  + a_1 \pqty*{\frac{2\mu}{ \omega}}
  + a_2 \pqty*{\frac{2\mu}{ \omega}}^{-1},
\end{equation}
with\footnote{This corrects a factor of \(3\) in~\cite{Hellerman:2023myh}.}
\begin{align}
   a_0 &= -0.0229\ldots, & a_1&= 0.3212\ldots, &  a_2 &= 2.6240\ldots 
\end{align}
\emph{This is not a truncation}:
the bulk contribution to the grand potential is a three-term polynomial\footnote{In the superfluid \ac{eft}~\cite{Son:2005rv,Beane:2025hon} to \ac{nlo}, 
\begin{align}
a_0 &= -\frac{5\pi^2}{128\sqrt{2}}\,c_0 = -\frac{1}{96\,\xi^{3/2}}, & a_1 &= -\frac{3\pi^2}{\sqrt{2}}\,c_1,
\end{align}
where the $c_i$ are low-energy constants that appear at \ac{nlo}, $\xi$ is the Bertsch parameter, and $a_2$ is fixed by \acs{nnlo} operators.}.
This does not mean that higher gap and density terms vanish; we are observing a property of their integral.

One might wonder if the termination of the expansion is an effect of the large-\(N\) approximation, but this is not the case.
To our knowledge, the cancellation of higher-order terms has not been previously observed, but it is already visible in the Son--Wingate \ac{eft}.
As discussed in~\cite{Hellerman:2021qzz}, the generic bulk term in the \ac{eft} has the form
\begin{equation}
  \mathcal{O}_{\text{bulk}}^{(m,n)}
  = (\partial_i X)^{2m} Z^n X^{d/2+1-(3m+2n)},
\end{equation}
where $\chi$ is the superfluid Goldstone, 
\begin{align}
  X &= \dot\chi - V(r) - \frac{1}{2}(\partial_i\chi)^2,
  &
  Z &= \Delta V - \frac{1}{d}(\Delta \chi)^2.
\end{align}
For \(V(r)=r^p\), this operator contributes to the grand potential as
\begin{multline}
	 \label{eq:Omega-m-n}
  \Omega_{m,n}
  \propto \int_0^1 d^d r\,\mathcal{O}_{\text{bulk}}^{(m,n)}\\ \propto \frac{1}{p}
  B \lrp*{
    \tfrac{d}{2}+2-3m-2n ,
    \tfrac{d+2m(p-1)+n(p-2)}{p}}   \mu^{d/2 + 1 + d/p - (1 + 2/p) (m+n)},
\end{multline}
where we have analytically continued the divergent integrals using Euler's beta function.
The simple observation is that \(B(x,y)\) has zeros if \(x+y\) is a non-positive integer, but neither \(x\) nor \(y\) is a non-positive integer.
This implies that some contributions vanish identically, independently of the unknown low-energy constants.
The case of the harmonic trap, \(p=2\), is particularly interesting.
If \(d\) is odd, \(\Omega_{m,n}\) vanishes whenever
\begin{equation}
  m+n\geq \frac{d+3}{2},
  \label{eq:vanishing_condition}
\end{equation}
which is precisely when the power of \(\mu\) is negative.
In other words, the entire negative-power part of the grand potential vanishes in odd dimensions.
If \(d=3\), only three terms survive, exactly as found above.

The same cancellation also occurs for anisotropic harmonic traps
in odd spatial dimensions. Keeping $d$ general, consider
\begin{equation}
 V(\mathbf r)
 =\frac12\left[
 \omega_\rho^2\sum_{i=1}^{d-1}r_i^2+\omega_z^2z^2
 \right].
\end{equation}
On the static saddle $\chi=\mu t$, the quantities entering
$\mathcal O_{\text{bulk}}^{(m,n)}$ become
\begin{align}
	X&=\mu-V,
 &
 Z&=\Delta V=(d-1)\omega_\rho^2+\omega_z^2.
\end{align}
We introduce the rescaled coordinates
\begin{align}
	s_i&=\frac{\omega_\rho r_i}{\sqrt{2\mu}},
 & i&=1,\ldots,d-1,
 &
 s_z&=\frac{\omega_z z}{\sqrt{2\mu}}.
\end{align}
The ellipsoidal cloud is then mapped to the unit ball $s<1$,
with
\begin{align}
	X&=\mu(1-s^2),
 &
 d^dr&=
 \frac{(2\mu)^{d/2}}{\omega_\rho^{d-1}\omega_z}
 s^{d-1}\dd{s} \dd{\Omega_{d-1}},
\end{align}
where $d\Omega_{d-1}$ is the angular measure on the unit
$(d-1)$-sphere. Taking the polar axis along $z$, the gradient
factor reads
\begin{equation}
 \partial_iX\,\partial_iX
 =2\mu s^2
 \left[
 \omega_\rho^2+
 (\omega_z^2-\omega_\rho^2)\cos^2\theta
 \right].
\end{equation}
Its angular integral can be evaluated explicitly:
\begin{multline}
	\int \dd{\Omega_{d-1}}
 \left[
 \omega_\rho^2+
 (\omega_z^2-\omega_\rho^2)\cos^2\theta
 \right]^m =\\
 2\pi^{(d-1)/2}
 \sum_{k=0}^{m}\binom{m}{k}
 \frac{\Gamma(k+\frac12)}{\Gamma(k+\frac d2)}
 \omega_\rho^{2(m-k)}
 (\omega_z^2-\omega_\rho^2)^k.
\end{multline}
The corresponding contribution to the grand
potential takes the form
\begin{multline}\label{eq:anisotropic-bulk-integral}
	\Omega_{m,n}\propto{}
 \frac{2^{m+d/2+1}\pi^{(d-1)/2}
       \big[(d-1)\omega_\rho^2+\omega_z^2\big]^n}
      {\omega_\rho^{d-1}\omega_z}\,
 \mu^{d+1-2(m+n)}
 \\
 \times
 \left[
 \sum_{k=0}^{m}\binom{m}{k}
 \frac{\Gamma(k+\frac12)}{\Gamma(k+\frac d2)}
 \omega_\rho^{2(m-k)}
 (\omega_z^2-\omega_\rho^2)^k
 \right]
 \\
 \times\int_0^1 \dd{s}
 s^{d-1+2m}
 (1-s^2)^{d/2+1-3m-2n}.
\end{multline}
All dependence on the trap frequencies appears in the prefactor.
The radial integral is identical to that of the isotropic
harmonic trap and can be expressed as
\begin{equation}
\int_0^1 ds\,
 s^{d-1+2m}(1-s^2)^{d/2+1-3m-2n}
 =
 \frac{1}{2} B \lrp*{
 \tfrac{d}{2} + 2 - 3m - 2n,\tfrac{d}{2} + m
 }.
\end{equation}
The cancellation therefore holds independently of the
anisotropy and of the low-energy constants.

\bigskip

The cancellation suggests that the harmonic trap may also be
exceptional at the quantum level. In particular, it is natural
to conjecture that the bulk grand potential receives no
additional logarithmic corrections beyond the universal one identified in~\cite{Hellerman:2021qzz}, which enters the grand potential at order \(\order*{\omega \log(\mu / \omega)}\).
A proof of this statement would require a more detailed
analysis of quantum fluctuations and renormalization,
which we leave for future work.

Even if true, this would not mean, however, that the full grand potential consists of only four terms.
As pointed out in Section~\ref{sec:perturbative}, the bulk expansion breaks down close to the edge, where contributions from different gradient orders must be resummed into an independent edge expansion (in the \ac{eft}, this is the effect of the edge operators in~\cite{Hellerman:2020eff}).
The vanishing of the bulk integrals does not affect the value of the edge contributions to \(\Omega\), which are generically nonvanishing, including for the harmonic potential.
We plan to return to this analysis in future work.

\subsection{$V(r)=r^4$}\label{Section:quartic-potential}

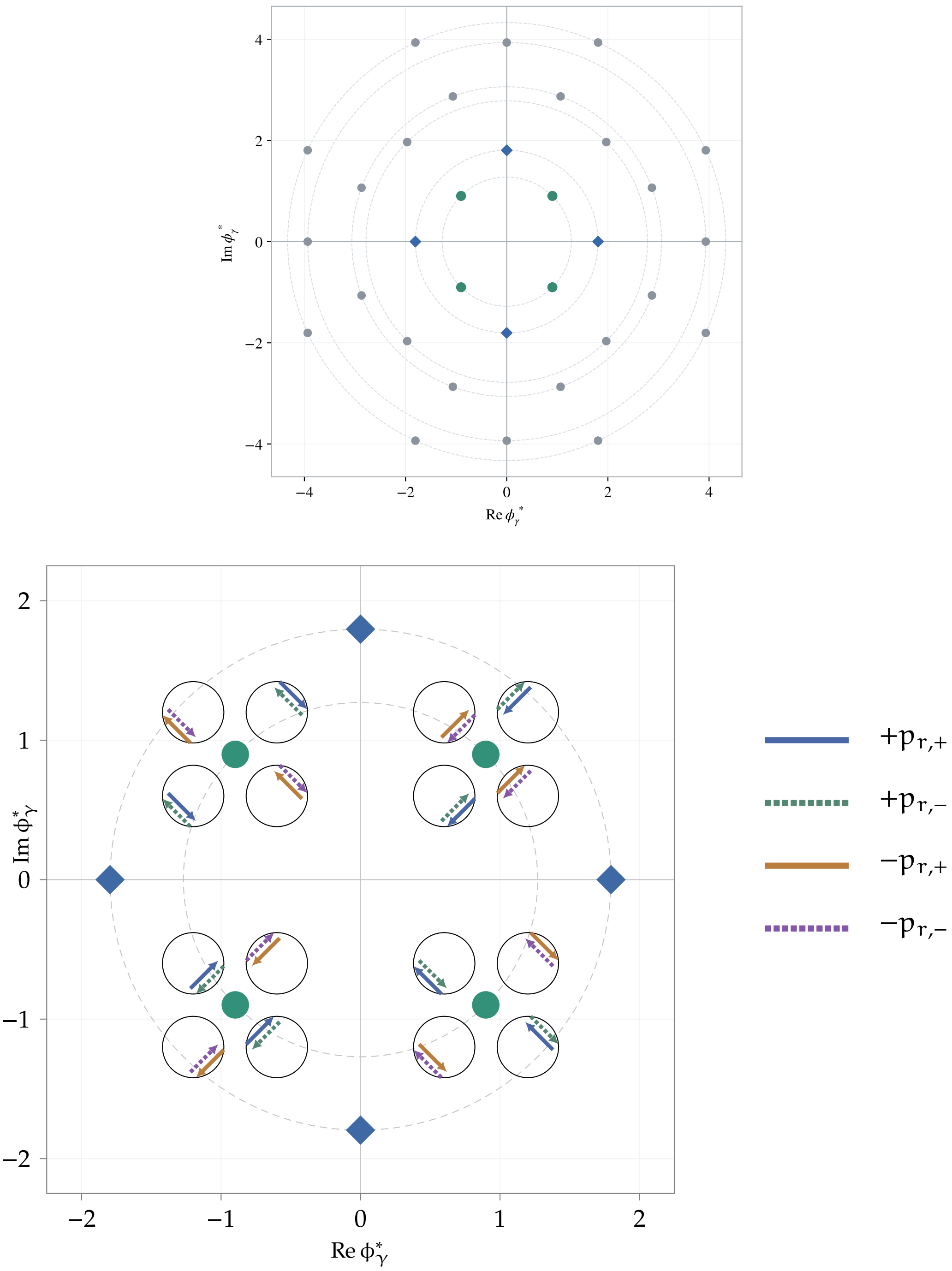
\begin{figure}
 \centering
 \includegraphics[width=.64\linewidth]{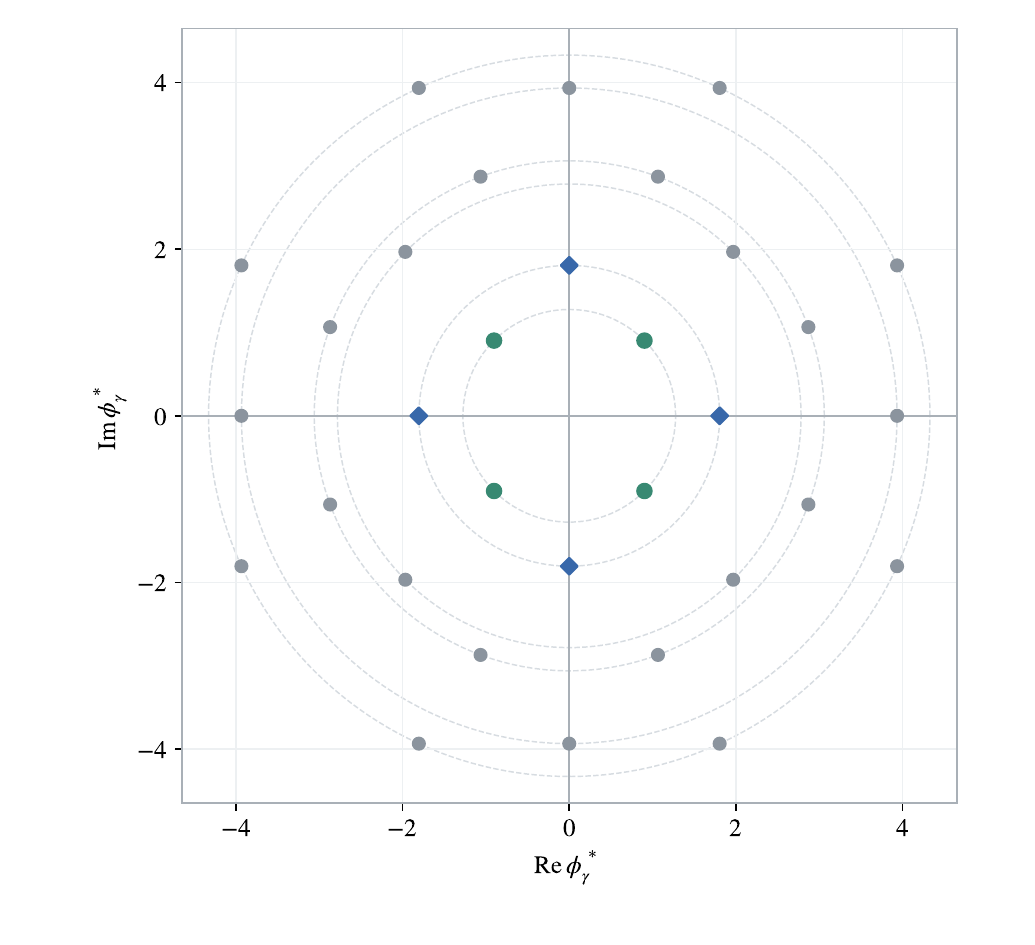}

   \begin{tikzpicture}[scale=2, font=\footnotesize]

    \def\rinner{1.27}
    \pgfmathsetmacro{\router}{1.27*sqrt(2)}
    \def\lim{2.25}

    \draw[step=1,gray!9,thin]
    (-\lim,-\lim) grid (\lim,\lim);

    \draw[gray!45,thin]
    (-\lim,0) -- (\lim,0);
    \draw[gray!45,thin]
    (0,-\lim) -- (0,\lim);

    \foreach \r in {\rinner,\router} {
      \draw[gray!45,thin,densely dashed]
      (0,0) circle[radius=\r];
    }

    \foreach \angle in {45,135,225,315} {
      \fill[pointgreen]
      (\angle:\rinner) circle[radius=2.8pt];
    }

    \foreach \angle in {0,90,180,270} {
    \path (\angle:\router) coordinate (P);
    \fill[diamondblue]
        ($(P)+(0,3pt)$) --
        ($(P)+(3pt,0)$) --
        ($(P)+(0,-3pt)$) --
        ($(P)+(-3pt,0)$) -- cycle;
}

    \foreach \x in {-2,-1,0,1,2} {
      \draw[gray]
      (\x,-\lim) -- ++(0,-0.05);
      \node[below,font=\small]
      at (\x,-\lim-0.05) {$\x$};
    }

    \foreach \y in {-2,-1,0,1,2} {
      \draw[gray]
      (-\lim,\y) -- ++(-0.05,0);
      \node[left,font=\small]
      at (-\lim-0.05,\y) {$\y$};
    }

    \draw[gray,thin]
    (-\lim,-\lim) rectangle (\lim,\lim);

    \node[below=16pt] at (0,-\lim)
    {$\Re \phi_\gamma^*$};

    \node[rotate=90,above=19pt] at (-\lim,0)
    {$\Im \phi_\gamma^*$};

    \circlearrows[0.2]{.9+.3}{-.9+.3}{0}{90}{mPm}{mPp}
    \circlearrows[0.2]{.9+.3}{-.9-.3}{0}{90}{pPp}{pPm}

    \circlearrows[0.2]{-.9+.3}{.9+.3}{0}{90}{pPm}{pPp}
    \circlearrows[0.2]{-.9+.3}{.9-.3}{0}{90}{mPp}{mPm}

    \circlearrows[0.2]{.9+.3}{.9+.3}{90}{180}{pPp}{pPm}
    \circlearrows[0.2]{.9+.3}{.9-.3}{90}{180}{mPm}{mPp}

    \circlearrows[0.2]{-.9+.3}{-.9+.3}{90}{180}{mPp}{mPm}
    \circlearrows[0.2]{-.9+.3}{-.9-.3}{90}{180}{pPm}{pPp}

    \circlearrows[0.2]{.9-.3}{-.9-.3}{180}{270}{mPp}{mPm}
    \circlearrows[0.2]{.9-.3}{-.9+.3}{180}{270}{pPm}{pPp}

    \circlearrows[0.2]{-.9-.3}{.9-.3}{180}{270}{pPp}{pPm}
    \circlearrows[0.2]{-.9-.3}{.9+.3}{180}{270}{mPm}{mPp}

    \circlearrows[0.2]{.9-.3}{.9-.3}{270}{360}{pPm}{pPp}
    \circlearrows[0.2]{.9-.3}{.9+.3}{270}{360}{mPp}{mPm}

    \circlearrows[0.2]{-.9-.3}{-.9-.3}{270}{360}{mPm}{mPp}
    \circlearrows[0.2]{-.9-.3}{-.9+.3}{270}{360}{pPp}{pPm}

    \begin{scope}[shift={(2.9,1.0)}]

      \foreach \col/\lab/\y in {
        pPp/{+p_{r,+}}/0,
        pPm/{+p_{r,-}}/-0.45,
        mPp/{-p_{r,+}}/-0.90,
        mPm/{-p_{r,-}}/-1.35
      } {
        \draw[\col,line width=2.5pt]
        (0,\y) -- (0.6,\y);

        \node[right,font=\large]
        at (0.75,\y) {$\lab$};
      }

    \end{scope}

  \end{tikzpicture}

 \caption{Top: the 32 distinct nonzero quartic \ac{wkb} action values.
 Green circles form the nearest adjacent-junction quartet; blue
 diamonds are the inner-sheet diameter actions. Gray points are the
 remaining sheet pairings. Endpoint geometry distinguishes the two
 nearest moduli.\\
Bottom: Trajectories for the smallest-action family. The turning points sit at \(r = \pm 1\) and \(r = \pm i\). The circles live in complex configuration space.}
 \label{fig:sec5-quartic-actions}
\end{figure}

The next example we consider is the quartic potential \(V=r^4\).
The junctions sit at the solutions of \(V(r)=1\), which are now the four roots of unity
\begin{align}
	 r_k &= e^{i \pi k/ 2}, & k&= 0,1,2,3. 
\end{align}
The shortest paths join two adjacent solutions and are shorter by a factor of \(\sqrt{2}\) than the diameter.
These paths lie in the complexified radial plane and have no direct counterpart among real classical trajectories.
The leading cycles correspond to inner-sheet exchanges (Andreev reflections) and have action
\begin{equation}
  \label{eq:quartic-minimal-action}
  \phi_\gamma^* = 1.2764\,e^{i\pi/4}.
\end{equation}
Rotations and reversals generate three further saddles of this type.
In total, there are 32 distinct nonzero \ac{wkb} action values (see Figure~\ref{fig:sec5-quartic-actions}).
Unlike the harmonic case, the quartic bulk coefficients do not exhibit the cancellation described above, and the expansion is asymptotic.
We expect its large-order behavior to be controlled by the minimal trajectories identified here.

The smallest-action family is complex, giving complex-conjugate candidate exponential factors.
Their phases predict a repeating \(+,+,-,-\) sign pattern in the perturbative coefficients.

\begin{figure}[t]
 \centering
 \includegraphics[width=\linewidth]{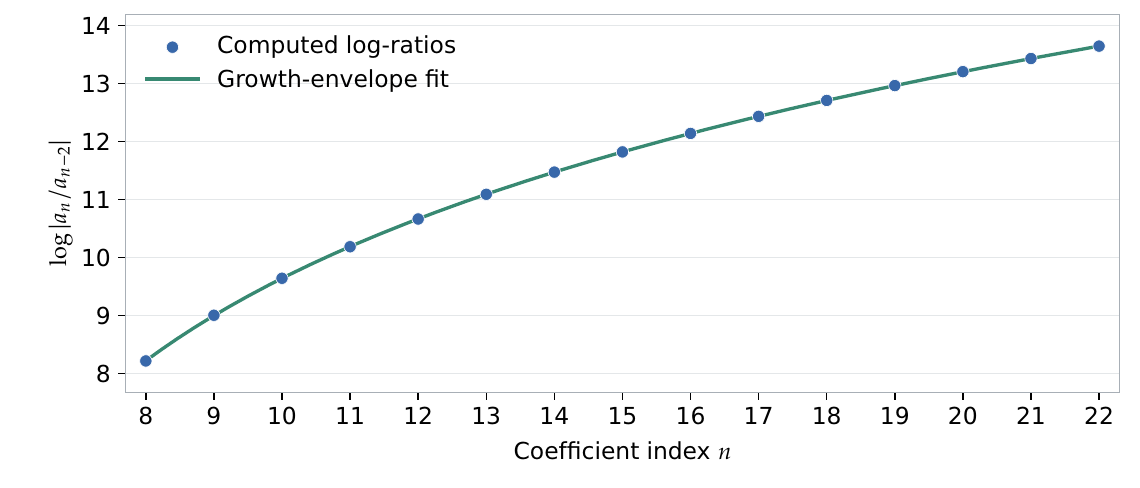}
 \caption{Quartic coefficient log-ratios $L_n=\log|a_n/a_{n-2}|$
for $8\leq n\leq22$, using coefficients computed up to order
$\epsilon^{44}$. Blue points show the computed log-ratios, and
the green curve is the growth-envelope fit over the displayed
window. This comparison tests the large-order growth; the absolute
values remove information about the oscillatory signs.}
 \label{fig:fit-quartic-growth}
\end{figure}

\begin{figure}[t]
  \centering
  \includegraphics[width=\linewidth]{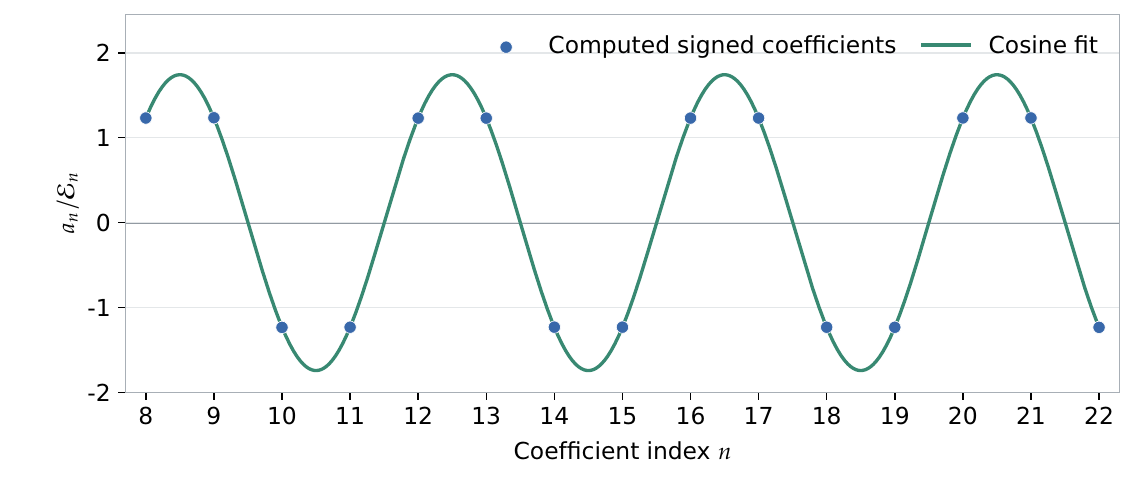}  
  \caption{Signed quartic coefficients after removing the fitted growth.
    Blue points are $\widetilde a_n$, with the actual signs
    retained, and the green line is the cosine obtained with
    $\theta$ free. The oscillation has an approximate four-coefficient period, with the repeating pattern $\set{+,+,-,-}$. }
  \label{fig:fit-quartic-signs}
\end{figure}

We have numerically computed the first 23 perturbative bulk coefficients $a_0,\dots,a_{22}$ up to order $\epsilon^{44}$. Explicitly,
\begin{equation}
  \frac{\Omega}{N\mu}
  \sim -0.0459 \frac{1}{\epsilon^3}
  + 0.6810 \frac{1}{\epsilon}
  + 16.9271 \epsilon + \cdots 
\end{equation}
We determine the growth parameters by fitting the logarithmic ratios $\log|a_n/a_{n-2}|$ for $n=8,\ldots,22$, which involve
the coefficients $a_6,\ldots,a_{22}$. Using the ansatz
(\ref{eq:perturbative-fit-ansatz}), we obtain
\footnote{The quoted errors are local least-squares errors at fixed fitting window and asymptotic ansatz;
robustness is assessed separately by varying both window endpoints, adding an inverse-order
correction, predicting excluded coefficients, and repeating both fitting stages after deletion of each
non zero input coefficient. Numerical precision checks and publication-rounding propagation are
negligible on these fitting scales over the tested coefficient ranges.}
\begin{align}
    A_{\text{fit}} &= 1.2754 \pm 0.0017,  &
  \beta_{\text{fit}} &= -2.529\pm 0.059, &
  c_{\text{fit}} &= 11.75\pm 0.69.
\end{align}
These results are shown in Figure~\ref{fig:fit-quartic-growth}. The quoted uncertainties are least-squares errors for the specified ansatz and the fixed fitting window $n=8,\dots,22$. We compare them with the action modulus from Eq.~\eqref{eq:quartic-minimal-action}:
\begin{equation}
  A_{\text{WKB}}= \abs*{\phi_\gamma^*} = 1.2764.
\end{equation}
The close agreement between the fitted action modulus and the geometric prediction supports the identification of the leading Andreev trajectory.
Holding the previously fitted growth parameters fixed, we fit $\widetilde a_n=a_n/\mathcal E_n$ to extract the angles \(\theta\), \(\delta\) and the amplitude $C$.
We obtain
\begin{align}
 \theta_{\text{fit}} &= (45.0009\pm0.0017)^\circ, &
 \delta_{\text{fit}} &= (45.031\pm 0.052)^\circ, &
 C_{\text{fit}} &= 1.81622\pm 0.00046,
 \label{eq:fit-quartic-phase-values}
\end{align}
to be compared to the \ac{wkb} prediction of \(\theta_{\text{WKB}}=  45^\circ\) (see Figure~\ref{fig:fit-quartic-signs}). The simultaneous agreement with the independently calculated \ac{wkb} action provides a nontrivial quantitative test of the proposed Andreev-cycle interpretation controlling the large-order behavior.

\begin{figure}
 \centering
 \includegraphics[width=.9\linewidth]{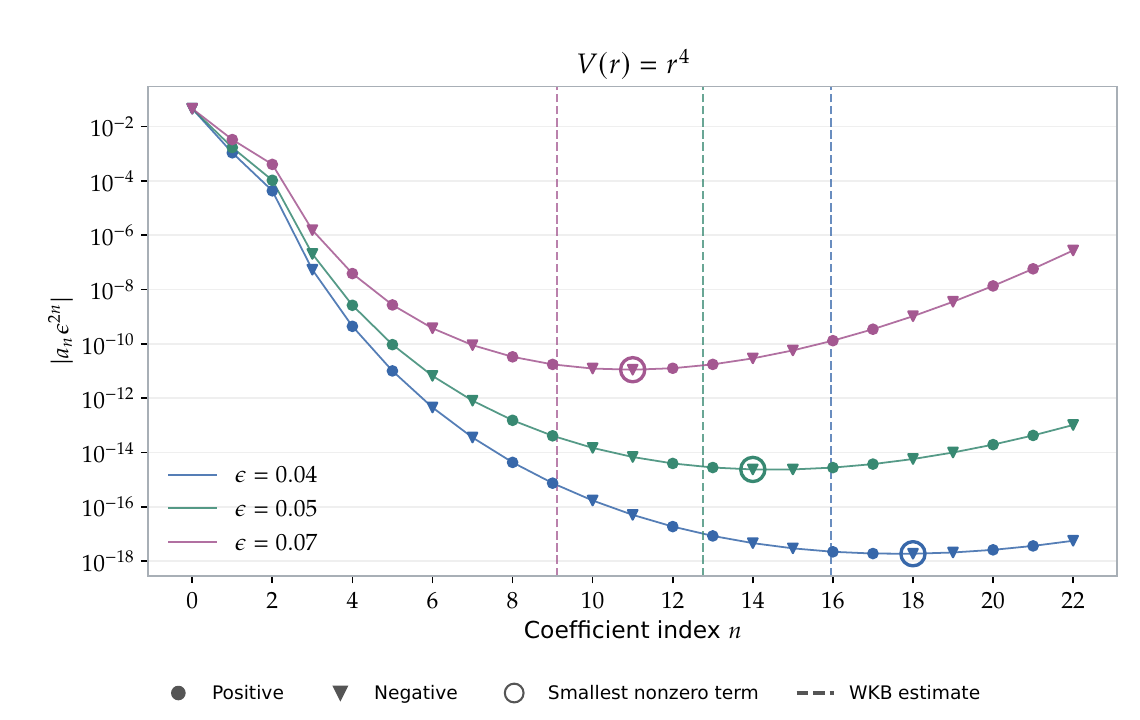}
 \caption{Quartic bulk term magnitudes $|a_n\epsilon^{2n}|$ through $n=22$,
 for $\epsilon=0.04,0.05,0.07$. Colors identify $\epsilon$; circles and
 triangles distinguish positive and negative terms. Open rings mark the
 smallest computed nonzero terms, at $n=18,14,11$, respectively, giving the optimal truncation.
 Dashed vertical lines of the same color show the leading \ac{wkb} estimates
 $|\phi_\gamma^*|/(2\epsilon)$.}
 \label{fig:wkb-quartic-terms}
\end{figure}

Having identified the action through both the semiclassical
geometry and the coefficient growth, we can predict how many orders of the bulk gradient expansion can be retained before its terms begin to grow. For this example, the identified action gives the leading estimate for the optimal truncation from Eq.~\eqref{equ:optimal-truncation-andreev}:
\begin{equation}
 N_{\text{opt}}\simeq
\frac{0.6382}{\epsilon}.
\end{equation}
As $\epsilon$ decreases, the minimum moves to higher order and its magnitude decreases. Thus more corrections can be retained before factorial growth takes over, while the least-term scale becomes smaller. 
Figure~\ref{fig:wkb-quartic-terms} compares the estimate with the computed perturbative terms for three values of $\epsilon$. The leading-order \ac{wkb} estimate lies slightly below the observed minimum. Corrections from the transport equation, associated with fluctuations around the trajectory, can shift the predicted value; these are not included in the present \ac{wkb} analysis.

\subsection{$V(r)=r^6$}
\begin{figure}
  \centering
  \includegraphics[width=.64\linewidth]{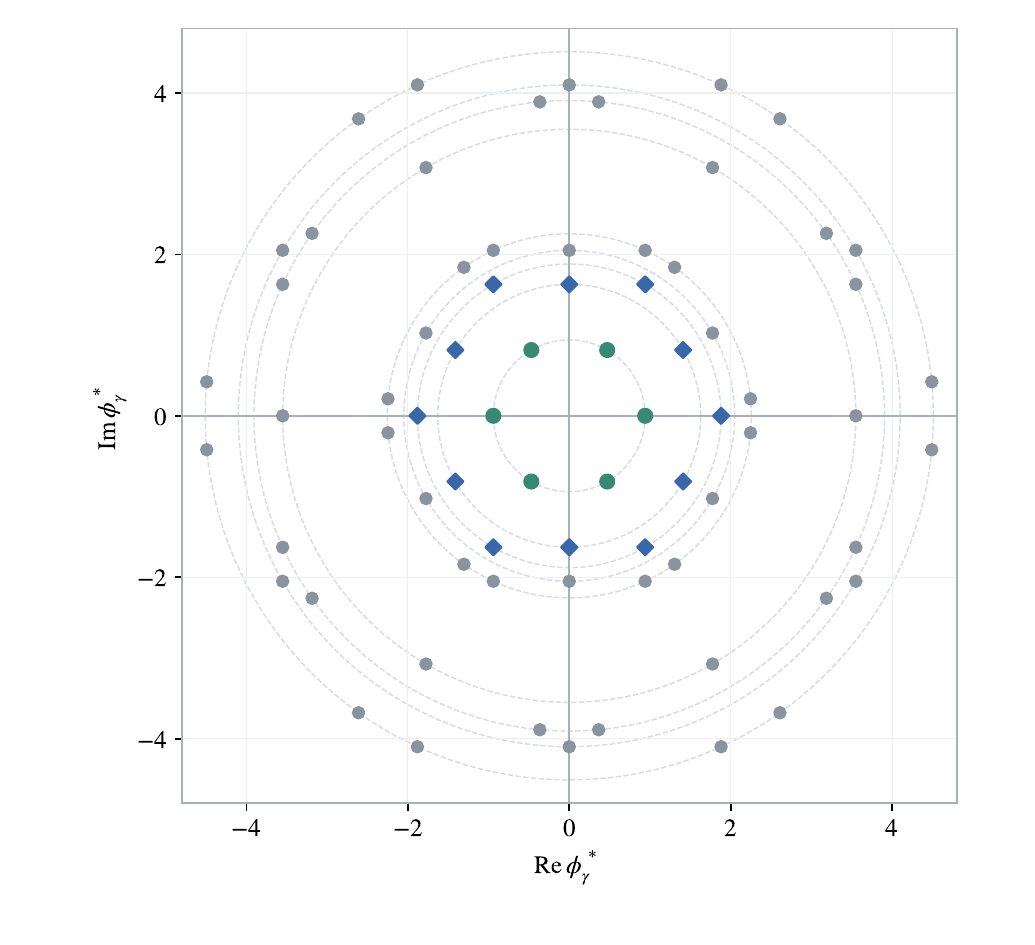}
  \caption{The 72 distinct nonzero sextic \ac{wkb} action values.
    Green circles form the nearest hexagon. Blue diamonds show
    inner-sheet exchanges across the longer chords; gray points show
    other sheet pairings. }
  \label{fig:sec5-sextic-actions}
\end{figure}

For the sextic potential, the junction points are $r_k = e^{ik\pi/3}$, $k=0,\ldots,5$.
The trajectories between adjacent turning points again give the nearest elementary family.
For $r_A=e^{2i\pi/3}$ and $r_B=e^{i\pi/3}$, the endpoint difference is $1$ and the action is real:
\begin{equation}
     \phi_\gamma^*
 =2  J_6
   \Im \sqrt{1+i y_0}
 =0.9405.
\end{equation}
Rotating both endpoints generates the hexagon
\begin{equation}
 \phi_\gamma^*\in
 \set{0.9405\,e^{ik\pi/3} |
 k=0,\ldots,5 }.
\end{equation}
The first terms of the grand potential expansion are
\begin{equation}
  \frac{\Omega}{N\mu}
  \sim -0.0612 \frac{1}{\epsilon^3}
  + 1.0432 \frac{1}{\epsilon}
  + 52.3134 \epsilon + \cdots
\end{equation}
The other endpoint separations are $\sqrt{3}$ and $2$, giving larger
inner-exchange moduli $\sqrt{3}$ and $2$ times the nearest one.
In particular a trajectory covering the diameter has twice the action of the minimal one.
These longer exchanges and the other sheet pairings are shown
in Figure~\ref{fig:sec5-sextic-actions}.

We apply the same procedure as in Section~\ref{Section:quartic-potential} to the sextic coefficients
$a_0,\ldots,a_{20}$, computed up to $\epsilon^{40}$.
Unlike in the quartic case, this sequence contains exact zeros:
\begin{align}
	a_{3m}&=0,& m&\geq1.
\end{align}
Every coefficient with index divisible by three vanishes. This all-orders property agrees with the spatial finite-part integration of \ac{eft} bulk terms in Eq.~\eqref{eq:Omega-m-n}. In particular, the term at order $\epsilon^3$ vanishes, and the next nonzero contribution is $1126.86 \epsilon^5$. 

\begin{figure}
  \centering
  \includegraphics[width=\linewidth]{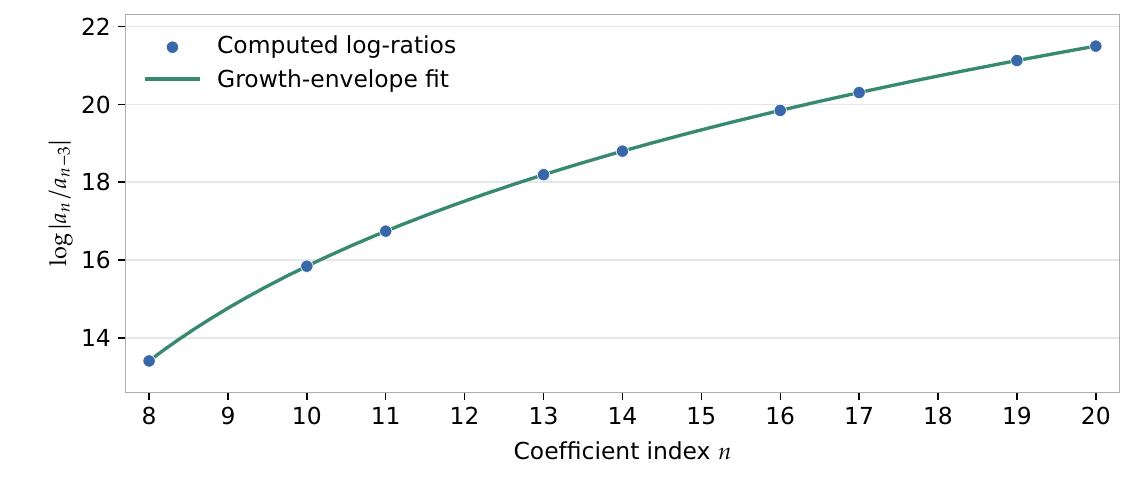}
  \caption{Sextic coefficient log-ratios $L_n=\log|a_n/a_{n-3}|$
for $8\leq n\leq20$, using coefficients computed through order
$\epsilon^{40}$. Indices divisible by three are omitted because
the corresponding coefficients vanish exactly. Blue points show
the computed log-ratios, and the green curve is the growth-envelope fit over the displayed window. This comparison tests the large-order growth; the absolute values remove information about the oscillatory
signs.}
  \label{fig:sextic_growth}
\end{figure}

Using the same procedure as above, we fit the nonzero logarithmic ratios $\log|a_n/a_{n-3}|$ (Figure~\ref{fig:sextic_growth}) and obtain
\begin{align}
  A_{\text{fit}}&=0.94066\pm 0.00023, &
\beta_{\text{fit}}&=-2.177\pm0.010, &
c_{\text{fit}}&=15.91\pm 0.12.
\end{align}
For comparison, the nearest \ac{wkb} family gives
\begin{equation}
  A_{\text{\ac{wkb}}}=0.94049\ldots.
\end{equation}
This is again in good agreement with the fitted result.
Since \(a_{3m}=0\) for \(m\geq1\), we extended the calculation to higher orders to obtain enough nonzero coefficients for the growth fit. The fitted action modulus agrees with the Andreev-cycle prediction within the estimated uncertainties.

\begin{figure}
  \centering
  \includegraphics[width=\linewidth]{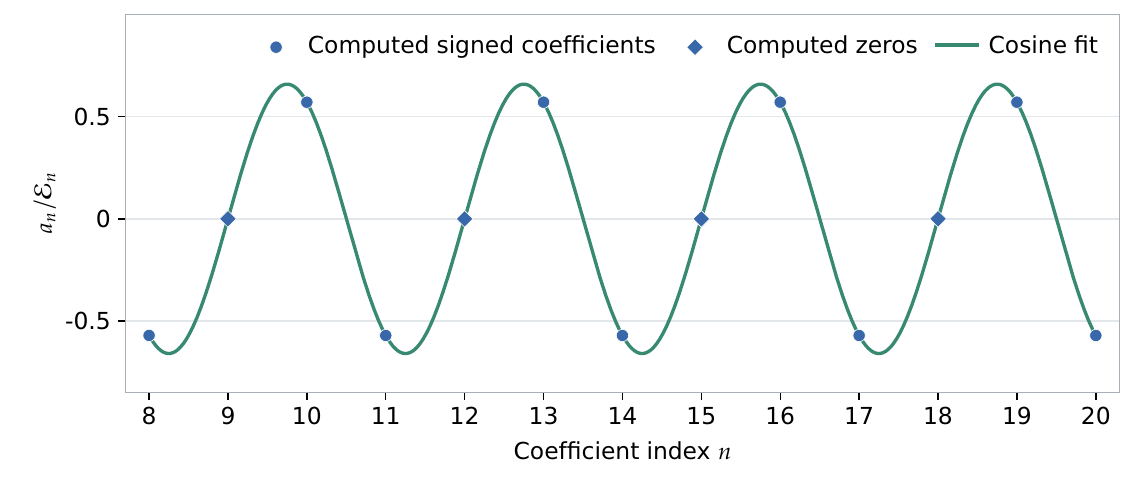}
    \caption{Signed sextic coefficients after removing the fitted growth.
    Blue points are $\widetilde a_n$, with the actual signs
    retained, and the green line is the cosine obtained with
    $\theta$ free.
    The oscillation has period three in the coefficient index, with the repeating sign pattern $\{+,-,0\}$, in agreement with $\theta_{\text{WKB}}=\pi/3$ and $\delta=\pi/2$.}
  \label{fig:sextic_sign}
\end{figure}

Holding the growth parameters fixed, we fit $ \widetilde a_n$ over
$n=8,\ldots,20$, including the exact zeros (Figure~\ref{fig:sextic_sign}).
The result is
\begin{align}
  \theta_{\text{fit}}&=(60.00037\pm 0.00045)^\circ, &
  \delta_{\text{fit}}&=(90.011\pm 0.013)^\circ, &
C_{\text{fit}}&=0.659481\pm 0.000035.
\end{align}
The fitted angles are to be compared with the complex \ac{wkb} images at
$\arg\phi_\gamma^*=\pm\pi/3$,
\begin{equation}
  \cos*(\frac{2\pi n}{3}-\frac{\pi}{2})
  =\sin*(\frac{2\pi n}{3}),
\end{equation}
which for $n$ integer exhibits the signs $+,-,0$, with period three.

The all-orders cancellation of the \(a_{3m}\) coefficients in the \ac{eft}, together with the numerical results, constrains the Borel structure and suggests that one of the Stokes parameters may vanish.
The actions of the closest trajectories sit at the vertices of a regular hexagon, so the leading asymptotic behavior of the \(a_n\) is
\begin{equation}
  a_n \sim \Gamma(2n + \beta) A^{-2n} \pqty*{ C_0 + C_1 e^{2 \pi i n /3 } + C_2 e^{- 2 \pi i n /3 }}  \, ,
\end{equation}
where the \(C_i\) parameterize the contributions associated with the corresponding actions.
The \(a_n\) are real, so \(C_2 = C_1^{*}\).
If \(a_{3m} = 0\) for all \(m \ge 1\), then the leading asymptotic behavior must obey
\begin{equation}
  C_0 + 2 \Re C_1 = 0 .
\end{equation}
Matching to the numerical coefficients we reproduce $\sin(2\pi n/3)$ for \(C_1\) being purely imaginary and \(C_0=0\).
If exact, this would mean that the trajectory with real action \(\phi_0^{*} = 0.9404\) has a vanishing Stokes parameter and does not contribute to the asymptotic expansion of the energy. In that case, it would define a semiclassical instanton without an active Borel singularity. The sextic problem may thus provide an explicit example in which the semiclassical spectral curve has an elementary real action, but the corresponding leading Borel amplitude vanishes for the integrated bulk observable.

\begin{figure}
 \centering
 \includegraphics[width=\linewidth]{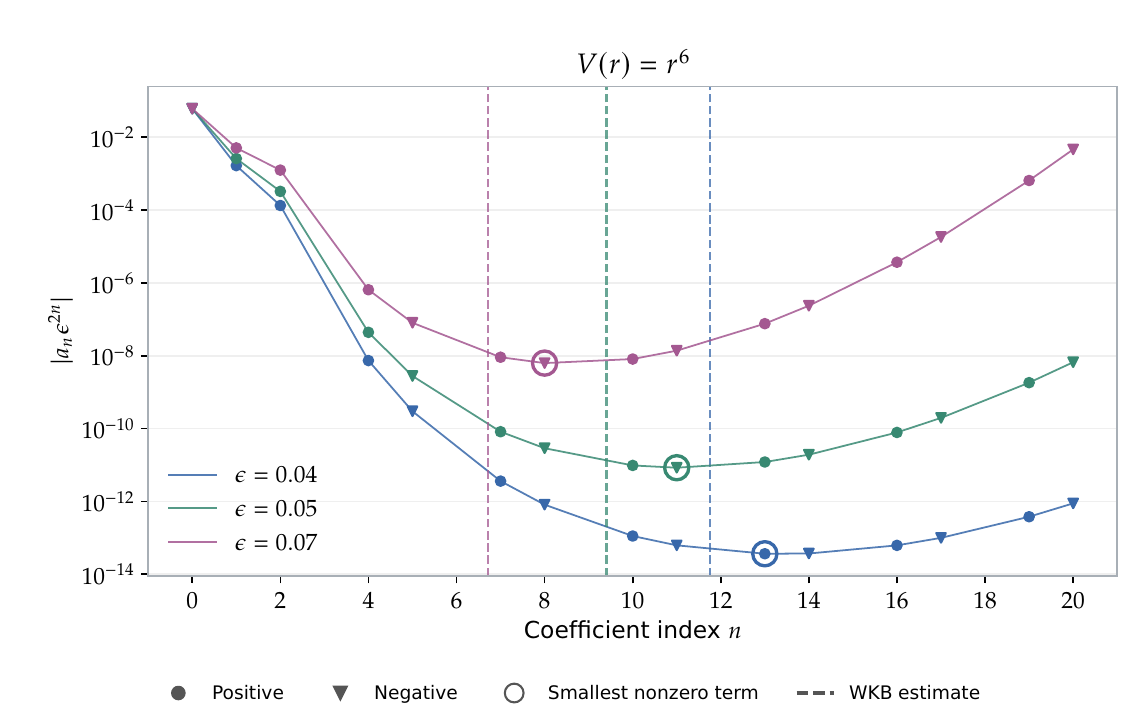}
 \caption{Sextic bulk term magnitudes through $n=20$, with the same
 conventions as Fig.~\ref{fig:wkb-quartic-terms}. The smallest nonzero terms (denoting the optimal truncation)
 occur at $n=13,11,8$ for $\epsilon=0.04,0.05,0.07$, respectively. Decreasing $\epsilon$ moves the
 minimum to higher order and reduces its magnitude.}
 \label{fig:wkb-sextic-terms}
\end{figure}

As done for the quartic potential, we can estimate the optimal truncation from this action:
\begin{equation}
 N_{\text{opt}}\simeq \frac{0.47025}{\epsilon}.
 \label{equ:OT-sextic}
\end{equation}
For \(\epsilon=0.04,\,0.05,\,0.07\), Eq.~\eqref{equ:OT-sextic} predicts \(N_{\text{opt}}\simeq11.8,\,9.4,\,6.7\), respectively, while the smallest terms occur at \(n=13,\,11,\,8\), as shown in Figure~\ref{fig:wkb-sextic-terms}. The differences are of order unity, consistent with the expected subleading corrections to the leading \ac{wkb} estimate. Comparing with the previous section, at fixed $\epsilon$, the flatter the potential in the center the lower the optimal truncation.

\FloatBarrier

\section{Free energy, conformal dimensions and fixed charge}
\label{sec:fixed-charge}

As is often the case in large-charge calculations, we have worked at fixed chemical potential \(\mu\), although physical observables are most naturally expressed in terms of the charge \(Q\).
In the \(Q\to\infty\) limit, the two descriptions are related by the Legendre transform in Eq.~\eqref{eq:Legendre-transform}.
However, the detailed large-order behavior of the fixed-charge perturbative coefficients is not identical to that in the grand potential \(\Omega\): the Legendre transform reorganizes the asymptotic expansion.
The leading nonperturbative saddle action, and hence the leading exponential factor, is unchanged by the Legendre transform, even though the perturbative tail is reorganized.

\begin{figure}
    \centering
    \includegraphics[width=.8\linewidth]{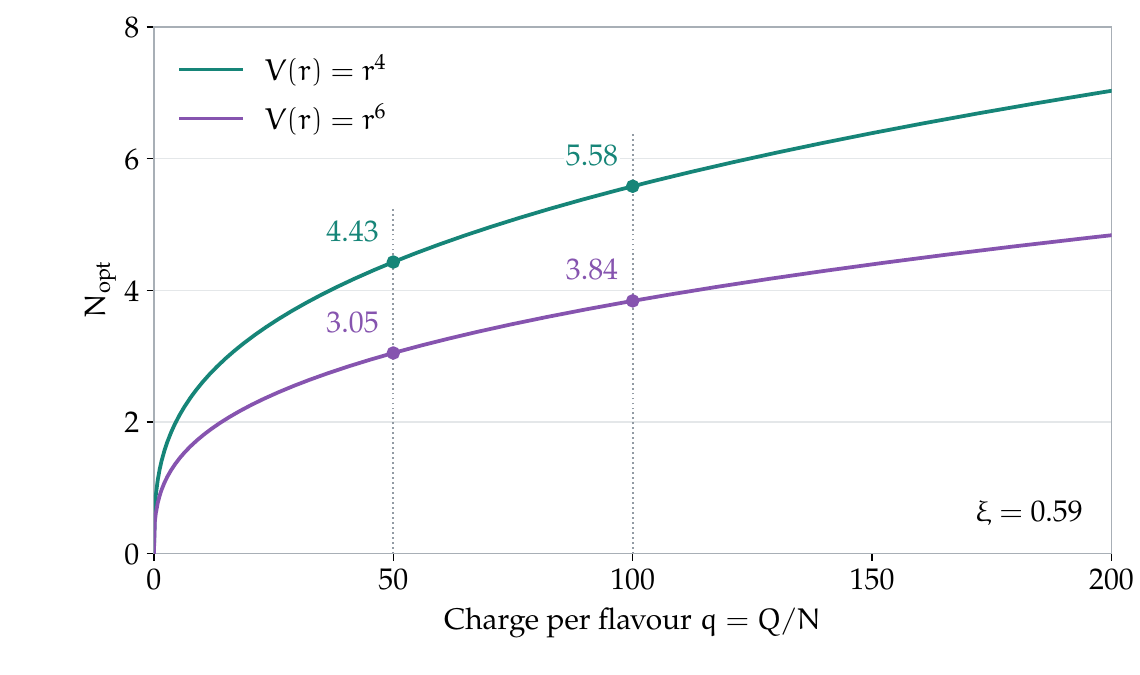}
    \caption{Leading \ac{wkb} estimates of the optimal truncation index \(N_{\mathrm{opt}}\) using~\eqref{Equ:OT-charge} versus charge per flavor \(q=Q/N\), for quartic (teal) and sextic (purple) traps, using \(\xi=0.59\). Both estimates grow as \(q^{1/3}\), with the sextic trap giving a smaller optimal index at fixed charge. Markers highlight \(q=50\) and \(q=100\).}
    \label{fig:optimal-truncations}
\end{figure}

In the case of the harmonic potential, we have found that the bulk contribution truncates at three terms.
The Legendre transform does not preserve this termination: the fixed-charge free energy is determined implicitly by the system
\begin{equation}
  \begin{dcases}
    \frac{F^{(2)}_{\text{bulk}}}{\omega N}
    = -\frac{3}{2}a_0 \epsilon^{-4}-\frac{1}{2} a_1 \epsilon^{-2}+\frac{a_2}{2}
    + \order{1/N},\\
    \frac{Q}{N}
    = -4a_0 \epsilon^{-3} - 2 a_1 \epsilon^{-1} +\order{1/N}.
  \end{dcases}
\end{equation}
The result still depends on the three coefficients already computed, although this dependence is less transparent. Alternatively, expanding \(F(Q)\) at large \(Q/N\) gives a convergent series whose coefficients depend only on \(a_0\), \(a_1\), and \(a_2\):
\begin{equation}
  \begin{aligned}
    \frac{F}{\omega N}
    \sim{}&
         \frac{3}{8\times 2^{2/3}|a_0|^{1/3}}\left(\frac{Q}{N}\right)^{4/3}
         +\frac{a_1}{4\times 2^{1/3}|a_0|^{2/3}}\left(\frac{Q}{N}\right)^{2/3}
         \\
       &+\frac{6a_0a_2-a_1^2}{12a_0} +\frac{a_1^3}{54\times 2^{2/3}|a_0|^{4/3}}
         \left(\frac{N}{Q}\right)^{2/3}  +\cdots \\
  \end{aligned}
\end{equation}
The nonperturbative structure we have found for the grand potential remains unchanged and the value of the saddle is obtained by taking the leading term of the expansion of $\mu(Q)$. Subleading terms would just contribute to the perturbative tail of each exponential. For \(V=r^p\), this gives
\begin{equation}
\epsilon_p
=
\frac{1}{\sqrt{\xi}}
\pqty*{
\frac{3\pi p}{4B(3/p,5/2)}\frac{Q}{N}
}^{-1/3},
\end{equation}
where \(\xi\) is the Bertsch parameter. For the harmonic oscillator,
\begin{equation}
\epsilon_2
=
\frac{1}{2\times 3^{1/3}\sqrt{\xi}}
\left(\frac{Q}{N}\right)^{-1/3},
\end{equation}
and the leading nonperturbative action is
\begin{equation}
\left. \frac{\phi_\gamma^*}{\epsilon}\right|_{\mathrm{harmonic}}
=
\frac{1.6221}{\epsilon}
\sim
3.60 \left(\frac{Q}{N}\right)^{1/3}.
\end{equation}
This is immediately generalized to \(V=r^p\):
\begin{equation}
 \eval*{\frac{\phi_\gamma^*}{\epsilon} }_{r^p}
=
\frac{2(6\pi)^{1/3}B(1/p,3/2)}{p^{2/3}B(3/p,5/2)^{1/3}}
\sin*(\frac{\pi}{p})
\sqrt{\xi}\,
\Im \sqrt{1+i y_0}
\left(\frac{Q}{N}\right)^{1/3}.
\end{equation}
In the general case, the asymptotic perturbative part in \(\Omega(\mu)\) is traded for an
asymptotic series in \(F(Q)\). For \(V=r^4\) and \(V=r^6\), the first terms are:
\begin{align}
  \frac{F_{\text{bulk}}^{(4)}}{N\Lambda_4}
  &\sim
    1.61169 \pqty*{\frac{Q}{N}}^{13/9}
    +2.98815 \pqty*{\frac{Q}{N}}^{7/9}
    +23.51903 \pqty*{\frac{Q}{N}}^{1/9}
    + \dots ,
  \\
  \frac{F_{\text{bulk}}^{(6)}}{N\Lambda_6}
  &\sim
    1.55587 \pqty*{\frac{Q}{N}}^{3/2}
    +4.28366 \pqty*{\frac{Q}{N}}^{5/6}
    +74.85081 \pqty*{\frac{Q}{N}}^{1/6}
    + \dots ,
\end{align}
where \(\Lambda_p=2^{-p/(p+2)}g_p^{2/(p+2)}\) and the potential is \(V(r) = g_p r^p\).
While the leading exponent depends on the potential, in both cases successive terms differ by a factor
of $(Q/N)^{-2/3}$. This is the reason why the large-order behavior, controlled by the action \(\phi^*(Q)\) above, has the same $Q/N$ dependence  for both potentials.
This leads to the optimal truncation (see Figure~\ref{fig:optimal-truncations}):
\begin{align}
  \eval*{ N_{\text{opt}}}_{r^4} &\approx 1.203 \left(\frac{Q}{N}\right)^{1/3}, \\
  \eval*{ N_{\text{opt}}}_{r^6} &\approx 0.827 \left(\frac{Q}{N}\right)^{1/3} .
  \label{Equ:OT-charge}
\end{align}

\FloatBarrier

\section{Conclusions}\label{sec:conclusions}

In this work, we have investigated the large-charge expansion of the unitary Fermi gas at leading order in large $N$, focusing on the perturbative bulk expansion of the grand potential and its nonperturbative semiclassical structure.
Using the heat-kernel formulation of the \ac{bdg} operator, we have combined a systematic gradient expansion with a semiclassical analysis of complex periodic trajectories.
This allows us to characterize the large-order behavior of the bulk expansion and determine its optimal truncation.

Our first main result concerns the case of a harmonic trapping potential which shows exceptional behavior.
We have shown that the integrated perturbative bulk expansion in $\mu$ of the grand potential terminates after three terms in three spatial dimensions.
Using the Son--Wingate \ac{eft}, we have demonstrated that this cancellation is independent of the low-energy constants and persists for anisotropic harmonic traps in any odd spatial dimension.
The termination is a property of the integrated bulk contribution and does not imply the absence of higher-order corrections to the complete grand potential, which also receives contributions from the droplet edge, which will be investigated separately.
The Legendre transform to fixed charge reorganizes the expansion, so that the corresponding expansion of the conformal dimension does not terminate but still depends only on three independent parameters.

Our second main result concerns anharmonic trapping potentials, for which the perturbative bulk expansion is instead asymptotic.
We have shown that the relevant semiclassical saddles are complex, periodic, zero-energy trajectories of the squared \ac{bdg} operator.
Their interpretation is closely related to Andreev reflection, analytically continued into complex phase space, where particle-like and hole-like branches of the spectral curve are exchanged.
For monomial potentials, the smallest candidate actions arise from cycles connecting adjacent complex turning points.
These actions determine the predicted positions of the leading Borel singularities and hence the factorial growth, oscillations, and optimal truncation of the perturbative coefficients.

We have tested these predictions for quartic and sextic trapping potentials by numerically computing the bulk gradient expansion to high orders.
The large-order behavior of the coefficients agrees remarkably well with the semiclassical predictions, both in the magnitude and in the phase of the leading complex actions.
For the quartic potential, the coefficients exhibit a repeating sign pattern of period four, while for the sextic potential every third coefficient vanishes exactly, in agreement with the \ac{eft} analysis.
The sextic case also suggests the possibility that a semiclassical trajectory with real action has a vanishing Stokes amplitude for the integrated bulk observable.
More generally, our results illustrate the importance of distinguishing the closed cycles of the semiclassical spectral curve from the elementary instantons of the theory and from the Borel singularities that actually contribute to the large-order behavior of a given observable.

The identification of the leading Borel singularities provides a quantitative prescription for the optimal truncation of the bulk gradient expansion, with the number of terms scaling as $N_{\mathrm{opt}}\simeq A/(2\epsilon)$, where $A$ is the modulus of the leading Borel singularity.
Since $\epsilon\sim (Q/N)^{-1/3}$, this establishes how the useful truncation order grows with charge.
The resulting estimates show that, even at relatively large charge, only a limited number of perturbative terms may improve the approximation before factorial growth sets in.

In our analysis, we have worked in the double-scaling limit $N\to\infty,\ Q\to\infty, Q/N$ fixed. Of course one wonders whether our results also apply to the physical case of one flavor. The truncation of the grand potential expansion in the harmonic case has been established from the \ac{eft} and is independent of $N$ which can only enter through the low-energy coefficients. The mechanism that generates the leading non-perturbative terms appears to be robust and independent of $N$. So we conjecture that any complex Andreev trajectories remain the dominant effect in controlling the $1/Q$ expansion. This means in turn that we expect the optimal truncation to always be at
\begin{equation}
	N_{\text{opt}} \sim c Q^{1/3} \sqrt{\xi},
\end{equation}
where $\xi$ is the Bertsch parameter and $c$ is a coefficient of order one. Note that even in the harmonic case, observables other than the grand potential will in general be asymptotic expansions in $Q$ and controlled in the same way.

\medskip
There are several interesting directions for future research.
In the present work, we have focused on the bulk contribution to the large-charge expansion.
The complementary analysis of the droplet edge, where the bulk gradient expansion breaks down and a separate boundary-layer description is required, will be the subject of a forthcoming paper.
Understanding the matching of the two regions will be essential for determining the asymptotic structure of the full grand potential.

On the nonperturbative side, it would be interesting to compute the fluctuation prefactors and Stokes coefficients associated with the complex Andreev trajectories, thereby clarifying the relation between semiclassical cycles and the Borel singularities of the perturbative expansion.

Finally, extending the semiclassical analysis to more general trapping potentials which more closely resemble experimental set-ups and to nonzero angular momentum may reveal additional families of complex trajectories governing the large-order behavior.

\section*{Acknowledgments}
\vspace{-1em}
\begin{small}\sffamily
  The authors would like to thank Subham Chowdhury and Félix Werner for useful discussions. 
  This work was supported by the Swiss National Science Foundation under grant number 200021\_219267.
  In addition, S.R.B is supported by the U.~S.~Department of Energy grant DE-FG02-97ER-41014 (UW Nuclear Theory).
  Claude Opus 5.5, GPT-5.6 Sol, and GPT-6 Astra were used in the preparation of this manuscript to assist with code drafting, debugging, and figure preparation. All AI-assisted outputs were reviewed and verified by the authors, who take full responsibility for the final content of the manuscript.
\end{small}

\appendix

\section{The cube and the ball}
\label{sec:CubeBall}

One source of technical difficulty in our study is the inhomogeneity of the configuration due to the confining potential. In this appendix, we consider the simpler cases of a system confined inside a cube or a ball. Most computations can be carried out explicitly, and the qualitative behavior is very similar to the case of the trap. The main difference is the lack of Andreev reflection, as we assume the gap \(\sigma\) to be constant in our approximation.

For \(V=0\) and \(\sigma\) constant, the \(B^2\) operator is proportional to the identity in the two-dimensional Nambu space:
\begin{equation}
B^2 = \pqty*{h^2+\sigma^2} \Id_2 \, .
\end{equation}
This does not mean that particles and holes do not mix, as the \ac{bdg} operator is not diagonal. However, since \([h,\sigma]=0\), no mechanism can convert a semiclassical particle-like branch into a hole-like one. There is no Andreev reflection.

From a technical point of view, this also simplifies the computation of the zeta function, which is now
\begin{equation}
   \zeta_{B^2}(s)=\frac{2}{\Gamma(s)}\int_0^\infty\frac{\dd{\tau}}{\tau} \tau^s e^{-\tau \sigma^2}\Tr*(e^{-\tau h^2}).
\end{equation}

\subsection{The cube}\label{sec:cube}

The first geometry that we consider is a cube of side \(L\) with Dirichlet \ac{bc}.
The corresponding heat kernel is
\begin{equation}
  K(t;\mu) = \Tr(e^{-\tau h^2})
  =
  \sum_{ n\in\setN^3}
  \exp\left[- \tau \left(
    \frac{\pi^2}{2L^2} \abs{n}^2 - \mu
  \right)^2\right].
\end{equation}
The sum is over \(\setN^3\). It is convenient to extend it over \(\setZ^3\) by introducing \(r\)-dimensional \(\Theta \) functions,
\begin{equation}
  \Theta_r(\tau ;\mu)
  =
  \sum_{n \in \setZ^r}
  \exp\left[-\tau \left(
    \frac{\pi^2}{2L^2} \abs{n}^2 - \mu
  \right)^2\right],
\end{equation}
and use inclusion--exclusion to rewrite
\begin{equation}
  K(\tau;\mu)
  =
  \frac{1}{8}\left[
    \Theta_3(\tau ;\mu)-3\Theta_2(\tau ;\mu)+3\Theta_1(\tau ;\mu)
    -e^{-\mu^2 \tau }
  \right],
\end{equation}
whose four terms are respectively the bulk, face, edge, and corner contributions.

We expect the Mellin integral to be dominated by small \(\tau\).
For this reason it is convenient to Poisson resum:
\begin{align}
  \Theta_r(\tau ;\mu)
  &=
  \left(\frac{L}{\pi}\right)^r
  \sum_{ m \in \setZ^r}
  \int_{\setR^r} \dd^r{p}
  \exp*[-\tau \left(\frac{p^2}{2}-\mu\right)^2
             -2i L m\cdot p ], & p_i &= \frac{\pi n_i}{L}.
\end{align}
This is already the structure we are looking for.
The sum over the \(m \in \setZ^r\) is a sum over windings, each controlled by the exponential of the length \(\abs{m}\).
The zero-winding sector gives the perturbative expansion, the others are exponential corrections equivalent to the semiclassical trajectories that we have found in the main part of the paper.

\subsubsection{Perturbative expansion}

It is useful to consider the family of functions introduced in Eq.~(3.36) of~\cite{Hellerman:2023myh},
\begin{equation}
  I_{m,n}(y)
  =
  -
  \frac{\Gamma(n-1/2)}{4\sqrt{2}\,\pi^{7/2}}
  \int_0^\infty\text{d}q\,q^{2+m}
  \left[(q^2-1)^2+y^2\right]^{1/2-n}
 .
\end{equation}
In terms of these functions, the zero-winding sector is
\begin{equation}
  \begin{aligned}
    Z_{\text{cube}}^{(0)}(s;y\mu,\mu)
    ={}&
    -\frac{8\pi^{3/2}L^3}{\Gamma(s)}
    \mu^{3/2-2s}I_{0,s+1/2}(y)
    \\
    &+\frac{6\sqrt{2}\,\pi^{5/2}L^2}{\Gamma(s)}
    \mu^{1-2s}I_{-1,s+1/2}(y)
    \\
    &-\frac{6\pi^{5/2}L}{\Gamma(s)}
    \mu^{1/2-2s}I_{-2,s+1/2}(y)
    -\frac{1}{8}\mu^{-2s}(1+y^2)^{-s}.
  \end{aligned}
\end{equation}
In~\cite{Hellerman:2023myh} only \(I_{0,0}\) appeared, consistently with
retaining only the leading bulk term.

To obtain the grand potential, we set \(s=-1/2\) and use
\(\Omega^{(0)}/N=-Z^{(0)}(-1/2)\), since
\(Z=\zeta_{B^2}/2\). There is a subtlety with the surface term:
for nonzero \(\sigma\), its Mellin integral has a pole at \(s=-1/2\),
\begin{equation}
  \int_0^\infty \dd{k}\,k
  \left[\left(\frac{k^2}{2}-\mu\right)^2+\sigma^2\right]^{-s}
  =\frac{\sigma^2}{4(s+1/2)}+\order{1}.
\end{equation}
This is a boundary divergence of the constant-gap model.
We subtract the
pole and define the finite part at a reference energy \(M\).
In terms of
the functions above, this amounts to replacing \(I_{-1,0}\) by
\begin{equation}
    I_{-1,0}^{\text{ren}}(y;M/\mu)
    =\frac{1}{8\sqrt{2}\,\pi^3} \bqty*{
      \sqrt{1+y^2}-\frac{y^2}{2}
      +y^2\log*(\frac{2(M/\mu)(1+\sqrt{1+y^2})}{y^2}) }.
\end{equation}
Equivalently, we take the finite part of \(M^{2s+1}Z^{(0)}(s)\) at
\(s=-1/2\).
The result for the cube is
\begin{multline}
	 \frac{\Omega_{\text{cube}}^{(0)}}{N} = -4\pi L^3\mu^{5/2} I_{0,0}(y)
   +3\sqrt{2}\,\pi^2L^2\mu^2 I_{-1,0}^{\text{ren}}(y;M/\mu)
    \\-3\pi^2L\mu^{3/2} I_{-2,0}(y)
    +\frac{\mu}{8} \pqty*{\sqrt{1+y^2}+ 1}.
\end{multline}
The finite surface contribution depends on the subtraction scale \(M\),
as expected for a boundary counterterm.
There is also an extra \(\mu /8 \) contribution from the reference term that for this geometry is necessary to obtain the normal-ordered grand potential.
These are features of the
constant-gap approximation with a Dirichlet wall; the gap profile at a
self-consistent edge is a different problem.

These are all the perturbative terms for the cube. As in the harmonic
potential case of Section~\ref{sec:vr=r2}, the expansion terminates,
although the surface term now contains a logarithm.
This can be viewed as a degenerate case of a divergent expansion associated with complex-conjugate saddles, with nearly all coefficients vanishing because the cube is flat (a \emph{Cheshire cat} effect).
For a curved domain, such as the three-dimensional ball \(B^3\), the infinite asymptotic series reappears, and its optimal truncation is controlled by the saddle actions.

\subsubsection{Non-perturbative terms.}

Now consider the winding sectors. Each of them contributes to the zeta proportionally to
\begin{equation}
Z^w_{\text{cube}} \propto \int_0^\infty \frac{\dd{\tau}}{\tau} \tau^s e^{-\sigma^2\tau}\int \dd^3{p} e^{-\tau(p^2/2-\mu)^2-2i L \abs{m} p} \, .
\end{equation}
In the double-scaling limit \(\sigma\to\infty\), \(\mu\to\infty\), \(y_0=\sigma/\mu=\order{1}\), we can solve the integral with a saddle point approximation after  rescaling
\begin{align}
 z&=\frac{p}{\sqrt{2\mu}} , & w  &=\tau \mu^{3/2} ,
\end{align}
and rewriting the exponent in the integrand as
\begin{equation}
-\tau\left[\sigma^2+\left(\frac{p^2}{2}-\mu\right)^2\right]-2i L \abs{m} p =-\sqrt{\mu}\left[w\left(y^2+(z^2-1)^2\right)+2i L \abs{m} \sqrt{2}\,z\right].
\end{equation}
In the limit \(\mu\to\infty\), we minimize with respect to \(w\) and \(z\) to get
\begin{equation}
\begin{cases}
 y^2+(z^2-1)^2=0,\\
 4wz(z^2-1)+2i \sqrt{2} L \abs{m}=0.
\end{cases}
\end{equation}
The first equation is nothing else than the condition that we had found in Section~\ref{subsec:non-perturbative} for a \(E=0\), \(L=0\) periodic complex trajectory:
\begin{equation}
\left(\frac{p_*^2}{2}-\mu\right)^2+\sigma^2=0.
\end{equation}
The action at the saddle then reduces to
\begin{equation}
e^{-2i L \abs{m} p_*} = e^{\pm 2iL|m|\sqrt{2(\mu\pm i\sigma)}},
\end{equation}
consistently with our Hamiltonian analysis.
The vectors \(m\in\setZ^r\) are billiard trajectories of length \(2 L \abs{m}\).
This gives a Balian--Bloch interpretation of the heat kernel~\cite{Balian:1970fw}.
The leading exponential contribution is the one with \(\abs{m}=1\).
Each of the three \(\Theta\) functions has such a term, so they all give the same exponential factor.
What changes is the leading \(\mu\) scaling of the perturbative part, which scales as \(\mu^{1/2-3s/2}\) for the bulk contribution, that is the dominant one.

\subsection{The ball}\label{sec:ball}

For the three-ball, the Laplacian eigenvalues are known explicitly in terms of
zeros of Bessel functions, but there is no comparably simple Poisson
resummation as we have used for the cube.
We instead start from the semiclassical Weyl density in
momentum space,
\begin{equation}
  g(k)
  =
  \frac{V}{2\pi^2}k^2
  -\frac{S}{8\pi}k
  +\frac{C}{12\pi^2}
  +\cdots.
\end{equation}
For a ball of radius \(R\), this becomes
\begin{equation}
  g(k)
  =
  \frac{2R^3}{3\pi}k^2
  -\frac{R^2}{2}k
  +\frac{2R}{3\pi}
  +\cdots.
\end{equation}
It immediately gives the perturbative part through
\begin{equation}
  k(t;\mu)
  \sim
  \int_0^\infty\text{d}k\,g(k)
  \exp*[-t\left(\frac{k^2}{2}-\mu\right)^2],
\end{equation}
namely
\begin{multline}
    Z_{B^3}^{(0)}(s;y\mu,\mu)
    -\frac{32\pi^{5/2}R^3}{3\Gamma(s)}
    \mu^{3/2-2s}I_{0,s+1/2}(y)
    +\frac{4\sqrt{2}\,\pi^{7/2}R^2}{\Gamma(s)}
    \mu^{1-2s}I_{-1,s+1/2}(y)
    \\
    -\frac{16\pi^{5/2}R}{3\Gamma(s)}
    \mu^{1/2-2s}I_{-2,s+1/2}(y)
    +\cdots.
\end{multline}
Setting \(s=-1/2\), with the same finite-part prescription for the
surface term, gives
\begin{equation}
    \frac{\Omega_{B^3}^{(0)}}{N}
    \sim-\frac{16\pi^2}{3}R^3\mu^{5/2}I_{0,0}(y)
    +2\sqrt{2}\,\pi^3R^2\mu^2 I_{-1,0}^{\text{ren}}(y;M/\mu)
    -\frac{8\pi^2}{3}R\mu^{3/2}I_{-2,0}(y)
    +\cdots.
\end{equation}
This is the direct analogue of the cube result, but we have only written the first terms in a divergent expansion, which is controlled by the nonperturbative exponentials.

The exponential corrections are associated with billiard trajectories.
In \(B^3\) the non-diameter polygon families are labelled by two positive integers, \((\nu,w)\) with
\(\nu>2w>0\), where \(\nu\) is the number of reflections and \(w\) is the
winding number around the center.  Their lengths are
\begin{equation}
  L_{\nu w}=2\nu R\sin\left(\frac{\pi w}{\nu}\right).
\end{equation}
The shortest orbit is the diameter, \((\nu,w)=(2,1)\), with
\(L_{21}=4R\).  %
However, an accidental symmetry of odd-dimensional spheres prevents the diameter orbit from controlling the Weyl asymptotic expansion of the heat kernel~\cite{Howls1999HighOO}.

In general, the Balian--Bloch trace formula gives the correction to the Weyl density:
\begin{equation}
    \delta g(k)
    =
    \sum_{\nu>2w>0}A_{\nu w}(k)
    \sin*(
      kL_{\nu w}-\frac{3\nu\pi}{2}-w\pi+\frac{3\pi}{4})
     -\sum_{r=1}^\infty A_r^{\text{d}}(k)\sin(4rRk),
\end{equation}
where the first term contains the polygons and the second the repetitions of
the diameter, with
\begin{align}
  A_{\nu w}(k)
  &=
  \frac{2R^{5/2}k^{3/2}}{\sqrt{\pi}}
  \sin\left(\frac{2\pi w}{\nu}\right)
  \sqrt{\frac{\sin(\pi w/\nu)}{\nu}}, &
   A_r^{\text{d}}(k)&=\frac{k R^2}{\pi r}.
\end{align}
The computation is then the same as for the cube.
A saddle point argument shows that every orbit \(\gamma_{v w}\) is associated to a universal exponential factor
\begin{equation}
  e^{-S_\gamma}
  = e^{\pm i L_{v w}\sqrt{2(\mu\pm i\sigma)}}.
\end{equation}

\section{Numerics}\label{sec:numerics}

Table~\ref{tab:quartic-sextic-coefficients} lists the perturbative coefficients of the leading-large-$N$ bulk grand potential%
\begin{equation}
  \frac{\Omega_{\text{bulk}}^{\text{FP}}}{N \mu}
  \sim\epsilon^{-3}\sum_{n\geq0}a_n\epsilon^{2n}   
\end{equation}
with
\begin{equation}
   \epsilon  = \frac{1}{\sqrt{2\mu} R_{\text{cl}}}
\end{equation}
for $V=g_p r^p$, $R_{\text{cl}}=( \mu/g_p)^{1/p}$.

Nonzero entries are rounded to 12 significant digits.
The sextic zeros at $n=3,6,9,\ldots$ are exact in the spatial finite-part prescription.
The entry $0^{\dagger}$ follows analytically, although the numerical sextic dataset ends at $\epsilon^{40}$. A dash denotes an unavailable coefficient.

\begin{table}[htbp]
  \centering
  \small
  \setlength{\tabcolsep}{9pt}
  \renewcommand{\arraystretch}{1.13}
  \begin{tabular}{lrr}
    \toprule
    $n$  & $a_n$ for $V\propto r^4$ & $a_n$ for $V\propto r^6$ \\
    \midrule
    $0$ & $-4.59627128141\times 10^{-2}$ & $-6.11999772871\times 10^{-2}$ \\
    $1$ & $6.81055250049\times 10^{-1}$ & $1.04321789881\times 10^{0}$ \\
    $2$ & $1.69271056314\times 10^{1}$ & $5.23134336698\times 10^{1}$ \\
    $3$ & $-1.30754796659\times 10^{1}$ & $0$ \\
    $4$ & $6.71829316097\times 10^{1}$ & $1.12685796013\times 10^{3}$ \\
    $5$ & $9.56172958247\times 10^{2}$ & $-2.84656345238\times 10^{4}$ \\
    $6$ & $-2.69542834431\times 10^{4}$ & $0$ \\
    $7$ & $-1.31541337965\times 10^{6}$ & $1.34272626282\times 10^{8}$ \\
    $8$ & $9.96991052755\times 10^{7}$ & $-1.89156598014\times 10^{10}$ \\
    $9$ & $1.07022625667\times 10^{10}$ & $0$ \\
    $10$ & $-1.53537381419\times 10^{12}$ & $1.01815883520\times 10^{15}$ \\
    $11$ & $-2.83858013279\times 10^{14}$ & $-3.52837794063\times 10^{17}$ \\
    $12$ & $6.57199317161\times 10^{16}$ & $0$ \\
    $13$ & $1.86037173169\times 10^{19}$ & $8.09890517393\times 10^{22}$ \\
    $14$ & $-6.31606527283\times 10^{21}$ & $-5.14013403068\times 10^{25}$ \\
    $15$ & $-2.53197347985\times 10^{24}$ & $0$ \\
    $16$ & $1.18288511614\times 10^{27}$ & $3.35751407902\times 10^{31}$ \\
    $17$ & $6.36844081948\times 10^{29}$ & $-3.37296596637\times 10^{34}$ \\
    $18$ & $-3.91336833711\times 10^{32}$ & $0$ \\
    $19$ & $-2.72184111970\times 10^{35}$ & $5.01235621983\times 10^{40}$ \\
    $20$ & $2.12703336892\times 10^{38}$ & $-7.29931840457\times 10^{43}$ \\
    $21$ & $1.85546266117\times 10^{41}$ & $0^{\dagger}$ \\
    $22$ & $-1.79625667582\times 10^{44}$ & $\text{---}$ \\
    \bottomrule
  \end{tabular}
  \caption{Perturbative coefficients of the leading-large-$N$, fixed-chemical-potential bulk grand potential expansion in powers of \(\epsilon\).}
  \label{tab:quartic-sextic-coefficients}
\end{table}

\FloatBarrier

\setstretch{1}

\printbibliography

\end{document}